\documentclass[nofootinbib,aps,preprint,superscriptaddress]{revtex4}
\usepackage{amsmath}
\usepackage{citesort}
\usepackage{epsfig}

\newcommand{\beqa}{\begin{eqnarray}}
\newcommand{\eeqa}{\end{eqnarray}}
\newcommand{\beq}{\begin{equation}}
\newcommand{\eeq}{\end{equation}}
\newcommand{\bal}{\begin{align}}
\newcommand{\eal}{\end{align}}

\def\gsim{\ \rlap{\raise 3pt \hbox{$>$}}{\lower 3pt \hbox{$\sim$}}\ }
\def\lsim{\ \rlap{\raise 3pt \hbox{$<$}}{\lower 3pt \hbox{$\sim$}}\ }

\newcommand{\fms}[1]{{#1}\!\!\!/}
\newcommand{\fmsl}[1]{{#1}\!\!\!\!/}
\newcommand{\half}{\frac{1}{2}}

\newcommand{\nn}{\frac{\fms{\overline{n}}}{2}} 
\newcommand{\nnn}{\frac{\fms{n}}{2}} 

\newcommand{\s}{\sigma_{\mu \nu}}
\newcommand{\n}{\overline{n}}
\newcommand{\bc}{\overline{c}}
\newcommand{\mP}{\mathcal{P}}
\newcommand{\w}{\omega}
\newcommand{\bw}{\bar{\omega}}
\newcommand{\SI}{\rm{SCET_I}}
\newcommand{\SII}{\rm{SCET_{II}}}
\newcommand{\T}{\mathcal{T}}

\newcommand{\vv}{\overline{v}}

\newcommand{\g}{\gamma}
\newcommand{\mS}{\mathcal{S}}
\newcommand{\C}{\mathcal{C}}

\begin{document}

\baselineskip 3.0ex 

\vspace*{18pt}

\title{Semi-inclusive hadronic $B$ decays in the endpoint region}

\def\addKorea{Department of Physics, Korea University, Seoul 136-701,
  Korea} 
\def\addPitt{Department of Physics and Astronomy, University of
  Pittsburgh, PA 15260, USA} 
\def\addcmu{Department of Physics, Carnegie Mellon University,
  Pittsburgh,  PA 15213, USA}
\def\addIJS{J.~Stefan Institute, Jamova 39, P.O. Box 3000, 1001
Ljubljana, Slovenia \vspace{1cm} }

\author{Junegone Chay}\email{chay@korea.ac.kr}\affiliation{\addKorea}
\author{Chul Kim}\email{chk30@pitt.edu}\affiliation{\addPitt} 
\author{Adam K. Leibovich}\email{akl2@pitt.edu}\affiliation{\addPitt} 
\author{Jure Zupan}\email{zupan@andrew.cmu.edu}
\affiliation{\addcmu}\affiliation{\addIJS}

\begin{abstract} \vspace*{18pt}
\baselineskip 3.0ex 
We consider in the soft-collinear effective theory 
semi-inclusive hadronic $B$ decays, $B\rightarrow XM$, in
which an energetic light meson $M$ near the endpoint recoils against
an inclusive jet $X$. We
focus on a subset of decays where the spectator quark from the $B$ meson
ends up in the jet. The branching ratios and direct CP asymmetries are
computed to next-to-leading order accuracy in $\alpha_s$ 
and to leading order in $1/m_b$. The contribution of charming penguins is
extensively discussed, and a method to extract it in semi-inclusive   
decays is suggested. Subleading  $1/m_b$ corrections
and $SU(3)$ breaking effects are discussed.
\end{abstract}

\maketitle

\section{Introduction} 
Semi-inclusive hadronic decays $B\to X M$ have received much less
attention over the years in contrast to the widely studied exclusive
two-body $B$ decays  
\cite{Cheng:2001nj,Atwood:1997de,Browder:1997yq,Soni:2005jj,Eilam:2002wu,Kagan:1997qn,He:1998se,Kim:2002gv,Calmet:1999ix}.  
As we will show in this paper,  semi-inclusive hadronic $B$ decays
in the endpoint region, where $M$ is an isolated energetic meson and
$X$ is a collinear jet of hadrons in the opposite direction, are 
theoretically simpler than the exclusive two-body $B$
decays in many respects, yet still address many of the questions that
had been debated in the context of the two-body $B$ decays.  Using the
soft-collinear effective theory 
(SCET) \cite{Bauer:2000ew,Bauer:2000yr,Bauer:2001ct,Bauer:2001yt}
predictions of semi-inclusive decays can be improved
systematically and lead to the following advantages. 
Firstly, larger data samples can be included by
considering inclusive jets with a variety of final-state particles
forming the collinear jet. Secondly, as 
in exclusive $B$ decays \cite{Chay:2003ju,Chay:2003zp}, the
four-quark operators in the weak Hamiltonian 
factorize at leading order in $1/m_b$ into a product of a
heavy-to-light current and a collinear current, with no strong
interactions between these two 
currents. Thirdly, the inclusive collinear jet is described by
the jet function, which is obtained by matching the full theory onto
$\mathrm{SCET}_{\mathrm{I}}$ at the scale $p_X^2 \sim m_b \Lambda$
with $\Lambda \sim \Lambda_{\mathrm{QCD}}$. Since the same
jet function appears at leading order in $B\to X_s\gamma$
or $B\to X l \overline{\nu}$ inclusive decays, many of the hadronic
uncertainties cancel by taking ratios. Finally, the contribution of
charming penguins can be implemented systematically using the
effective theory.  Studying $B\rightarrow XM$ decays can thus offer a
theoretical handle to probe 
nonperturbative effects of charming penguins.  

In order to see these advantages clearly, we consider the decays
$B\rightarrow XM$ in which the spectator quark of the $B$ meson goes
to the inclusive jet. It is straightforward to treat other decay modes
without this constraint \cite{work}, but would involve more
calculation including 
spectator interactions, and we do not discuss it further here. 
The decay rate for $B\to X M$ at leading
order in $1/m_b$ can then be schematically written as
\begin{equation} 
\frac{d\Gamma}{dE_M} \sim \big(|\T \otimes \phi_M|^2 \big)\cdot
\big(\mathcal{J} \otimes f\big), 
\label{decayr}
\end{equation} 
where $\T$ is a collection of hard coefficients obtained in matching
full QCD onto SCET$_{\rm I}$ and $\mathcal{J}$ is the discontinuity of
the jet function describing the fluctuations of order $m_b \Lambda$ in
the forward scattering amplitude of the heavy-to-light currents.   
$\phi_M$ and $f$ are the light-cone distribution amplitude (LCDA) for
the light meson $M$ and the $B$-meson shape function, respectively.
The $\otimes$ sign implies the appropriate convolution.
The convolution $\mathcal{J} \otimes f$ in Eq.~\eqref{decayr} is
universal in the sense that the same convolution appears in
$B\to X_s \gamma$ and $B\to X_u l\overline{\nu}$
decays \cite{Bauer:2000ew,Bauer:2003pi}. 
Therefore, if we take the ratios of the decay rates for $B\to X
M$ and, say, the decay rate for $B\to X_s \gamma$, this
convolution cancels out and the only surviving nonperturbative
parameters are the LCDAs. 

Another interesting but complicated problem  common to
two-body $B$ decays and $B\rightarrow X M$ decays in the endpoint
region is the contribution of intermediate charming penguins,
which can be of nonperturbative nature \cite{Ciuchini}. There has been
a disagreement on how to treat this contribution 
between the recent SCET analysis of the two-body $B$ decays
\cite{Bauer:2004tj,Williamson:2006hb} and the approach of QCD
factorization \cite{Beneke:1999br}.  
The question is whether or not the long-distance effects of
charming penguins are of leading order in $1/m_b$. 
Long-distance contributions arise when intermediate charm
quarks lie in the non-relativistic QCD (NRQCD) regime with small
relative velocity $v^*$. These contributions are of the form
$\alpha_s(2 m_c) f(2m_c/m_b) v^*$ \cite{Bauer:2005wb}, where
$f(2m_c/m_b)$ is a factor which accounts for the phase space in which 
the charm quarks have small relative velocities. In 
QCD factorization \cite{Beneke:1999br,Beneke:2004bn}, the claim is that the
phase space suppression near the threshold region is strong
enough so that the nonperturbative contributions are  
subleading. On the other hand, Bauer et
al.~\cite{Bauer:2004tj,Bauer:2005wb} argue that since $2 m_c/m_b$ is
of order unity so is  $f(2m_c/m_b)$, and the nonperturbative
contribution of charming 
penguins can be of leading order. In this paper, we suggest how to
resolve the issue of  charming penguins in $B\rightarrow X M$
decays. If the nonperturbative contributions of charming penguins are
really suppressed, then the decay rates at leading order in $1/m_b$
are completely determined in terms of the perturbatively calculable
hard kernels convoluted with LCDA, once normalized to the $B\to
X_{s}\gamma$ rate. If nonperturbative charming penguins are not
suppressed, they will show up experimentally as a sizable deviation
from purely perturbative predictions, which we will discuss in detail.  

The paper is organized as follows: In Section \ref{kinematics} we
describe the kinematics for $B \rightarrow M X$ decays.
In Section \ref{matching}, we set up the operator basis for the decays
$B \rightarrow MX$ and compute the radiative corrections at
next-to-leading order (NLO) to derive the renormalization group
equations for the operators. In Section~\ref{fac}, we present a
factorized form for the semi-inclusive $B$ decays in the endpoint
region. Section \ref{charmpengsec} is devoted to the contribution of
charming penguins, considering two possible scenarios in which the
charm quark is regarded as either hard-collinear or heavy. The
contribution of charming penguins in the heavy quark limit $m_b,
m_c\rightarrow \infty$ with $m_c/m_b$ fixed is considered in detail.
In Section~\ref{beyondLO}, we discuss the corrections to the leading
order prediction, including $SU(3)$ breaking effects. In Section
\ref{pheno}, we perform the phenomenological analysis of $B
\rightarrow MX$ decays and predict the decay rates and CP
asymmetries. The method to extract the effect of charming penguins is
also discussed. We conclude in Section~\ref{conclusion}. In Appendix A
we present the Wilson coefficients for 
the operators at NLO in $\mathrm{SCET}_{\rm I}$. In Appendix B the
detailed analysis of charming penguins in the heavy quark limit is
discussed.

\section{Kinematics}\label{kinematics}
Using SCET, solid predictions can be made for hadronic semi-inclusive 
$B \to M X$ decays in the endpoint region. In the rest
frame of the $B$ meson, the outgoing energetic meson $M$ with $p_M^2
\sim \Lambda^2$ moves in the $\overline{n}^{\mu}$ direction, while the
inclusive 
hard-collinear jet with $p_X^2 \sim \Lambda m_b$ is in the
$n^{\mu}$ direction, where $n^2=\overline{n}^2 =0$, $n \cdot
\overline{n} =2$. We can choose the reference frame in which
the transverse momentum of $M$ is zero.  The momenta $p_M^{\mu}$
and $p_X^{\mu}$ can be
written in terms of the light-cone coordinates $p^{\mu} =(\overline{n}
\cdot p, n\cdot p, p_\perp)$ as
\begin{eqnarray}
p_M &=& \left(0,n\cdot p_M, \vec 0\right)+ {\cal
  O}(\Lambda^2/m_B), \nonumber \\
p_X &=& \left(m_B, m_B - n\cdot p_M,\vec 0\right) + {\cal
  O}(\Lambda^2/m_B), 
\end{eqnarray}
with $p_B^{\mu} = m_B v^{\mu} =
p_M^{\mu} +p_X^{\mu}$, where $2v^{\mu} = n^{\mu}
+\overline{n}^{\mu}$. We consider the endpoint region in which $m_B -
n\cdot p_M \sim \Lambda$, so that $p_X^2 \sim m_B \Lambda$. 

At the quark level, the $b$ quark has momentum $p_b^\mu= m_b v^\mu +
l^\mu$, where $l^\mu$ is the residual momentum of order $\Lambda_{\rm 
  QCD}$. The $b$ quark decays to a quark--antiquark pair moving in the
$\overline{n}$ direction which hadronizes into the meson $M$, and
another quark with momentum $p_J^{\mu}$ moving in the $n$ direction,
which combines with a spectator antiquark to form the outgoing jet
$X$. The momentum $p_J^{\mu}$ can be written as (dropping terms of
order $\Lambda^2/m_b$) 
\begin{equation}
p_J^\mu = m_b v^\mu + l^{\mu} -  p_M^\mu  =   
m_b \frac{n^{\mu}}{2} + m_b(1-x_M) \frac{\overline{n}^\mu}2 + l^{\mu}
= m_b \frac{n^{\mu}}{2} +k^{\mu},
\end{equation}
where $x_M  = n\cdot p_M /m_b = 2E_M/m_b$. In the endpoint region,
$1-x_M \sim \Lambda/m_b$. Since the invariant mass squared $p_J^2$ of
the jet is  time-like, the range of the residual momentum $k^{\mu}$ in
$p_J^{\mu}$ is $0 \le n\cdot k \le n\cdot p_X$. Since
the residual momentum of the heavy quark $n\cdot l$ is smaller than 
$\overline{\Lambda} = m_B -m_b$, the region of $n\cdot l$, which has
support for the $B$-meson shape function, is 
\begin{equation} 
-m_b (1-x_M) \le n\cdot l \le \overline{\Lambda}.
\label{ndotl}
\end{equation}

\section{Matching and evolution in $\SI$}
\label{matching}  
\begin{figure}[b]
\begin{center}
\epsfig{file=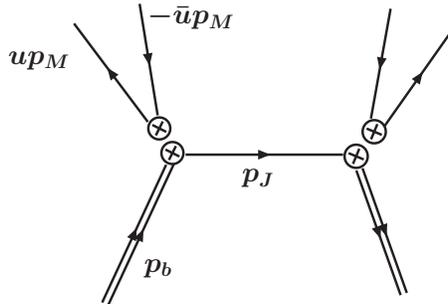, width=6cm}
\end{center}
\vspace{-1.0cm}
\caption{\baselineskip 3.0ex 
Tree-level diagram for the forward scattering of the heavy-to-light
currents in $\SI$ whose discontinuity gives the semi-inclusive
hadronic $B$ decay rates in the endpoint region.  
The double lines denote a heavy quark, the intermediate line is  
the hard-collinear quark in the $n$ direction with $p_J^2=m_b^2
(1-x_M)$, while the upward collinear quarks move in the $\overline{n}$
direction. } 
\label{kin}
\end{figure}

We employ a two-step matching in computing and evolving the hard
coefficients. First we construct the operators for the decays in $\SI$
by integrating out degrees of freedom of order $m_b$. The Wilson
coefficients of the operators are obtained by matching full QCD
onto $\SI$. The decay rates of the semi-inclusive $B$ decays are
obtained from the forward scattering amplitude of the 
time-ordered product of the heavy-to-light currents, as shown in
Fig.~\ref{kin}. In the next step, we match
$\mathrm{SCET}_{\mathrm{I}}$ onto $\mathrm{SCET}_{\mathrm{II}}$ by 
integrating out the degrees of freedom with  $p^2 \sim m_b
\Lambda$. As a result, the jet function is obtained, the discontinuity
of which contributes to the semi-inclusive hadronic $B$ decay rates.     

The effective weak Hamiltonian in full QCD for hadronic $B$ decays is
given as 
\begin{equation} 
H_{W} = \frac{G_F}{\sqrt{2}} 
\Biggl[ \sum_{p=u,c}  \lambda_p^{(q)} \Bigl(C_1 O_1^p + C_2 O_2^p
\Bigr) - \lambda_t^{(q)} \Bigl( \sum_{i=3}^{10} C_i O_i + C_g O_g +
C_{\gamma} O_{\gamma}\Bigr) \Biggr], 
\label{fullw}
\end{equation} 
where the operators are 
\begin{alignat}{2} 
O_1^p &= (\overline{p}b )_{V-A}
(\overline{q}p)_{V-A},\qquad\qquad\qquad\quad&  
O_2^p &= (\overline{p}_{\beta}b_{\alpha} )_{V-A} 
(\overline{q}_{\alpha}p_{\beta} )_{V-A},  \label{weakhamiltonian}\\  
O_{3,5} &= (\overline{q}b )_{V-A} \sum_{q'} (\overline{q'}q' )_{V\mp A},   
& O_{4,6} &= (\overline{q}_{\beta}b_{\alpha} )_{V-A} \sum_{q'} 
(\overline{q'}_{\alpha}q'_{\beta} )_{V\mp A}, \nonumber \\  
O_{7,9} &= (\overline{q}b )_{V-A} \sum_{q'} \frac{3e_{q'}}{2}
(\overline{q'}q')_{V\pm A}, &  
O_{8,10} &= (\overline{q}_{\beta}b_{\alpha})_{V-A} \sum_{q'}
\frac{3e_{q'}}{2} (\overline{q'}_{\alpha}q'_{\beta} )_{V\pm A},
\nonumber \\   
O_{\gamma} &= - \frac{em_b}{8\pi^2} \overline{q} 
\sigma^{\mu\nu} F_{\mu\nu} (1+\gamma_5) b, & 
O_g &= - \frac{gm_b}{8\pi^2} \overline{q} \sigma^{\mu\nu} 
 G^a_{\mu\nu}T^a (1+\gamma_5) b. \nonumber 
\end{alignat}
Here $\lambda_p^{(q)}= V_{pb}V_{pq}^*$ is the CKM factor and $V\pm A =
\gamma^{\mu} (1\pm \gamma_5)$. The summation over $q^{\prime}$
includes $u$, $d$, $s$, $c$ and $b$ quarks. Operators with $q=d$
($q=s$) describe the $\Delta S=0$ ($\Delta S=1$) effective weak
Hamiltonian. 

The effective Hamiltonian in $\SI$ at leading order (LO) in $1/m_b$ is
(with charm quarks integrated out, nonperturbative charm contributions
will be discussed in Section \ref{charmpengsec})
\cite{Chay:2003ju,Bauer:2004tj}    
\begin{equation} 
H_{I} = \frac{2G_F}{\sqrt{2}} \sum_{p=u,c} \lambda_p^{(q)}
\sum_{i=1}^6 \C_{i}^p \otimes  {\cal O}_i,
\label{hscet}
\end{equation}  
where the relevant four-quark operators are
\begin{alignat}{2}
\label{siop}
{\cal O}_1 &= \big[\overline{u}_n \fms{\overline{n}} P_L 
Y_n^{\dagger} b_v \big] \big[\overline{q}_{\n}  \fms{n} P_L u_{\n}
\big]_u, \qquad\qquad& 
{\cal O}_{2,3} &= \big[\overline{q}_n  \fms{\overline{n}} P_L
Y_n^{\dagger} b_v\big] 
\big[\overline{u}_{\n} \fms{n} P_{L,R} u_{\n}\big]_u,  \\ 
\label{sfquark}
{\cal O}_4 &= \sum_{q'} \big[\overline{q'}_n \fms{\overline{n}} P_L
Y_n^{\dagger} b_v\big] \big[\overline{q}_{\n} \fms{n} P_L
q'_{\n}\big]_u, & {\cal O}_{5,6} &= \sum_{q'} \big[\overline{q}_n
\fms{\overline{n}} P_L Y_n^{\dagger} b_v\big] \big[\overline{q'}_{\n}
\fms{n} P_{L,R} q'_{\n}\big]_u. \nonumber
\end{alignat}
The summation over $q'$ includes $u$, $d$ and $s$ quarks and $P_{L,R}
= (1\mp \gamma_5)/2$. In Eq.~(\ref{hscet}), $\otimes$ denotes the
convolution  
\begin{equation} 
\C_{i}^p\otimes {\cal O}_i = \int_0^1 du~ \C_{i}^p (u)  {\cal O}_i (u),
\end{equation} 
and the subscript $u$ in Eq.~\eqref{siop} refers to the variable in a
$\delta$ function, which is defined as 
\begin{equation} 
\big[ \overline{q}_{\n} \fms{n} P_L q_{\n} \big]_u\equiv\Bigl[
\overline{q}_{\n}\; \delta\! \Big(u-\frac{n\cdot
  \mathcal{P}^{\dagger}}{2E_M}\Big) \fms{n} P_L q_{\n} \Bigr]. 
\end{equation} 
The  $q_n$ and $q_{\bar{n}}$ are the gauge-invariant quark fields
\begin{equation}
  q_n = W_n^{\dagger} \xi_n^{(q)}, \ q_{\bar{n}} = W_{\bar{n}}^{\dagger}
  \xi_{\bar{n}}^{(q)}, \label{qn}
\end{equation}
given in terms of the collinear fermion fields $\xi_n^{(q)}$,
$\xi_{\bar{n}}^{(q)}$ of  flavor $q$ and the collinear  
Wilson lines $W_n$, $W_{\bar{n}}$ in the $n$ and $\overline{n}$
directions, respectively.  The ultrasoft (usoft)   
Wilson line in the $n$ direction, $Y_n$, is obtained after redefining
the collinear fields to decouple collinear and usoft degrees of freedom
\cite{Bauer:2001yt}. 

There are also color-octet operators corresponding to the operators in
Eq.~\eqref{hscet}, e.g., 
\begin{equation} 
\overline{\mathcal{O}}_1 (u) = \big[ \overline{u}_n Y_n^{\dagger} Y_{\n}
\fms{\overline{n}} 
P_L T^a  Y_{\n}^{\dagger} b_v \big] \big[ \overline{q}_{\n} \fms{n}
P_L T^a u_{\n}\big]_u,
\label{octet}
\end{equation}
but the matrix elements of the octet operators between 
hadronic states vanish and are therefore not relevant here. 
The Wilson coefficients $\C_i^p(u)$ in Eq.~\eqref{hscet} encode
physics at the hard scale $m_b$ and are perturbatively calculable in
powers of $\alpha_s(m_b)$.  They are known at NLO in $\alpha_s$
\cite{Beneke:1999br,Chay:2003ju} and are listed in Appendix 
\ref{appA}. Note that the Wilson coefficients $\C_i^p(u)$ exhibit
nonzero strong phases at NLO from integrating out the 
intermediate on-shell quarks.

\begin{figure}[b]
\begin{center}
\epsfig{file=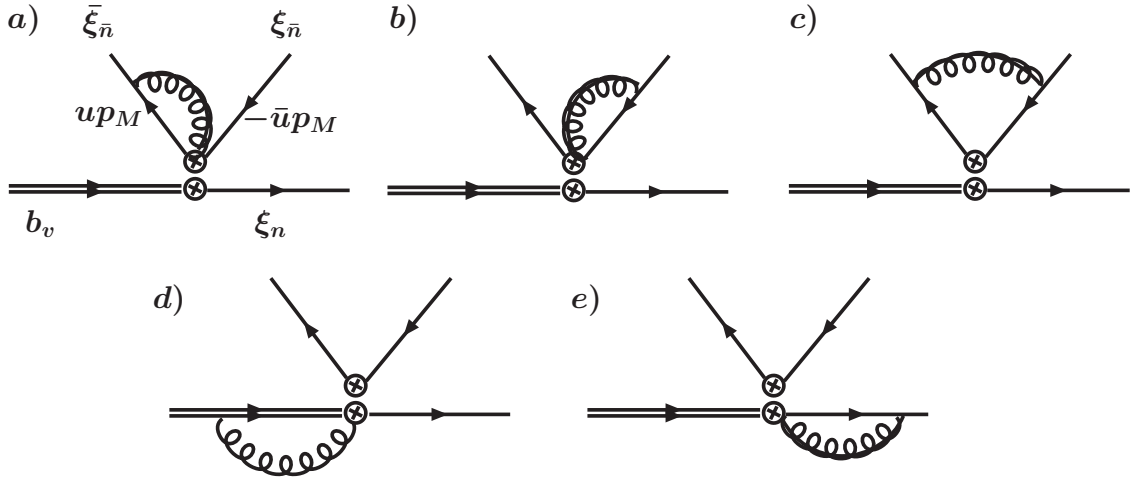, width=15cm}
\end{center}
\caption{\baselineskip 3.0ex 
Feynman diagrams at order $\alpha_s$ for the four-quark operators 
in $\SI$. Note that there is no strong interaction between the two
currents.}   
\label{fquark}
\end{figure}

In matching $\SI$ onto $\SII$ at $\mu_0\sim \sqrt{m_b\Lambda}$, the
operators in the Hamiltonian in Eq.~\eqref{hscet} are first evolved
down from $m_b$ to $\mu_0$ using the renormalization group (RG)
equation in $\SI$. The operators in Eq.~\eqref{hscet} factor into
a heavy-to-light current $J_H^{\mu}$ and a collinear current
$J_C^{\mu}$ as\footnote{${\cal O}_4$ is a sum over a product of
currents.  When   considering the spectator quark going
into the jet, only   one of the terms will contribute.} 
\begin{equation} 
{\cal O} (u,\mu)= \Bigl[ \overline{q}_n \Gamma_H  Y_{n}^{\dagger} b_v
\Bigr] \Bigl[\overline{q}_{\n} \Gamma_C q_{\n} \Bigr]_u = J_H (\mu)
\cdot  J_C (u,\mu), 
\label{geop}
\end{equation}
where $\Gamma_{H,C}$ are the Dirac structure in each current. There
are no strong interactions between the two currents 
to all orders in $\alpha_s$ in $\SI$. At order $\alpha_s$, the
radiative corrections in Fig.~\ref{fquark} show explicitly that this
is true. As a result, the operator $\cal O$ is
multiplicatively renormalized, and there is no mixing
between color-singlet and color-octet operators due to factorization.
The renormalized operator ${\cal O}_R$ and the
bare operator ${\cal O}_B$ are related by
\begin{equation} 
{\cal O}_R (u,\mu)
= \int dv Z^{-1}(u,v,\mu) {\cal O}_B (v,\mu)
=\int dv Z_H^{-1}(\mu) Z_C^{-1}(u,v,\mu) {\cal O}_B (v,\mu), 
\label{renormal}
\end{equation} 
where the counterterm $Z$ is a product of the counterterms
$Z_H$ and $Z_C$ from the radiative corrections of $J_H^{\mu}$ and
$J_C^{\mu}$. This leads to the RG equation
\beq
\begin{split}  \label{RGE}
\mu\frac{d}{d\mu} 
{\cal O} (u,\mu) &= \Bigl(\mu\frac{d}{d\mu} J_H (\mu) \Bigr) J_C
(u,\mu) +J_H(\mu)  \mu\frac{d}{d\mu} J_C (u,\mu)  \\
&= -\gamma_H J_H(\mu) J_C(u,\mu) - J_H(\mu)
\int dv \gamma_C (u,v,\mu) J_C(v,\mu), 
\end{split} 
\eeq
whereas the RG equation for the Wilson coefficients is written as
\begin{equation} 
\mu\frac{d}{d\mu} \C_i (u,\mu)
= \int_0^1 dv \Bigl[\gamma_H (\mu)
\delta(u-v) +\gamma_C (v,u,\mu) \Bigr] \C_i (v,\mu).
\label{CRGE}
\end{equation} 
At next-to-leading logarithm (NLL),  the anomalous dimensions for
$J_H$ and $J_C$ are given by
\begin{eqnarray} 
\gamma_H (\mu) &=& -\frac{\alpha_s C_F}{\pi}  \ln\frac{\mu}{m_b}
-\frac{\alpha_s C_F}{2\pi} \Bigl(\frac{5}{2} 
+ \frac{\alpha_s }{\pi} B\ln\frac{\mu}{m_b} \Bigr),\\
\gamma_C (u,v,\mu) &=& -\frac{\alpha_s C_F}{2\pi} 
\Biggl[\frac{3}{2}\delta(u-v)+
2\frac{u}{v}\Bigl(1+\frac{1}{(v-u)_+}\Bigr)\theta(v-u) + \Bigl(u,v
\leftrightarrow \bar u, \bar v \Bigr) \Biggl],  
\end{eqnarray}   
where $C_F =(N^2-1)/(2N)$ with $N$ the number of colors. The subscript
`+' denotes the plus distribution, and $\bar u =1-u$. 
The one-loop result for $\gamma_H$ in SCET was first obtained in
Ref.~\cite{Bauer:2000ew}, while the part of the two-loop result
containing $\alpha_s^2\ln(\mu/m_b)$ needed at NLL accuracy has not yet
been calculated in SCET. Extracting it from the full QCD calculation
\cite{Korchemsky:1994jb}, one gets $B = N(67/18-\pi^2/6)-5n_f /9$,
where $n_f$ is the number of flavors. The one-loop result for
$\gamma_C$ can be taken from the full QCD hard kernel calculations
\cite{Lepage:1980fj,Chernyak:1983ej}, which agree with the
determination in SCET \cite{Fleming:2004rk}.

At one loop, Eq.~\eqref{CRGE} can be written as 
\begin{equation} 
\mu\frac{d}{d\mu} \C_i (u,\mu) = \left[ \gamma_H (\mu) -
  3\frac{\alpha_s C_F}{2\pi} \right]  \C_i(u,\mu) -\frac{\alpha_s
  C_F}{2\pi} \frac{1}{u\bar u}\int_0^1 dv V(u,v) \C_i(v,\mu),   
\label{DRGE}
\end{equation} 
where $V(u,v)$ is the Brodsky-Lepage kernel \cite{Lepage:1980fj} 
\begin{equation} 
V(u,v) = 2\Bigl[\bar u v\Bigl(1+\frac{1}{(u-v)_+}\Bigr)
\theta(u-v) +\Bigl(u,v \leftrightarrow \bar u, \bar v\Bigr) \Bigr].  
\end{equation} 
The eigenfunctions of Eq.~(\ref{DRGE}) are given by the Gegenbauer
polynomials $C_n^{3/2}(2u-1)$, which satisfy 
\begin{equation} 
\int^1_0 dv V(u,v) C_n^{3/2}(2v-1) = e_n u \bar u C_n^{3/2}(2u-1),
\end{equation}
with the eigenvalues 
\beq
e_n = -4 +\frac{2}{(n+1)(n+2)}-4\sum_{k=2}^{n+1}\frac{1}{k},
\eeq
since the four-quark operators in $\SI$ are partially governed by the
light-cone conformal symmetry with the highest weight of the conformal
spin $j=2+n$ \cite{Braun:2003rp}. 

We can now expand the Wilson coefficients in terms of the Gegenbauer
polynomials,
\begin{equation} 
\C_i(u,\mu) = \sum_{n=0} C_n^{3/2}(2u-1)~A_i^n (\mu),
\label{sumD}
\end{equation}  
a virtue of which is that $A_i^n$ with different $n$ do not mix to one
loop. The solution of the RG equations for $A_i^n$,
\begin{equation} 
\mu\frac{d}{d\mu} A_i^n (\mu) = \Bigl[\gamma_H (\mu) - 3\frac{\alpha_s
  C_F}{2\pi} -\frac{\alpha_s 
  C_F}{2\pi} e_n \Bigr] A_i^n(\mu),
\end{equation}
yield the Wilson coefficient at the scale $\mu$ to NLL order 
\beq 
\C_i (u,\mu) = \sum_n C_n^{3/2}(2u-1)~A_i^n(m_b) \exp [I_n(\mu,m_b)],
\eeq 
with
\begin{eqnarray} 
I_n (\mu,m_b) &=& \frac{4\pi}{\alpha_s(m_b)}\frac{C_F}{\beta_0^2}
\Bigl[1-\frac{1}{z}-\ln z\Bigr] + \frac{\beta_1}{\beta_0^3}C_F 
\Bigl[-1 +z -\ln z +\half \ln^2z \Bigr] \nonumber \\
\label{in}
&+&\frac{C_F}{\beta_0}\Bigr(\frac{11}{2} +e_n \Bigr) \ln z 
+ \frac{2B}{\beta_0^2}C_F \Bigl[1-z+\ln z\Bigr], 
\end{eqnarray} 
where $z =\alpha_s(\mu)/\alpha_s(m_b)$. The coefficients of the QCD 
$\beta$ function are $\beta_0=(11-2n_f)/3$, and $\beta_1=34N^2/3-10N
n_f/3 -2C_F n_f$.  From the orthogonality of the Gegenbauer
polynomials the coefficients $A_i^n(m_b)$ are
\begin{equation} \label{ain}
A_i^n (m_b) = \frac{4(2n+3)}{(n+1)(n+2)}\int^1_0 du ~u(1-u)\C_i (u,m_b)
C_n^{3/2}(2u-1).
\end{equation} 
At NLL, only LO values of the Wilson coefficients at $\mu=m_b$ are
needed. Since these are independent of the momentum fraction $u$, we
have $A_i^n (m_b)=\C_{i,\mathrm{LO}}(m_b) \delta_{n0}$.

\section{Semi-inclusive decay rates}\label{fac} 
The decay amplitudes for $B\to XM$, in which a spectator quark of the
$B$ meson ends up in the jet $X$, are schematically 
\beq
\langle X M | H_I |B \rangle 
= \frac{2G_F}{\sqrt{2}} \int^1_0 du ~T_M^{(q)} (u,\mu) 
~\langle M |\big[\overline{q}'_{\n} \fms{n} P_L
q''_{\n} \big]_u|0 \rangle 
\langle X | \overline{q}_n \fms{\overline{n}} P_L
Y_n^{\dagger} b_v | B \rangle, \label{amp}
\eeq
where the hard kernel $T_M^{(q)}$ is given by the sum of the products
of the CKM factors $\lambda_p^{(q)}=V_{pb}V_{pq}^*$ and
the Wilson coefficients $\C_{i}^p$. Here $q$ denotes the flavor of the
outgoing quark in the heavy-to-light current. The matrix
elements for the meson $M$ are related to the LCDA by
\begin{eqnarray}
\langle P | \big[\overline{q}'_{\n} \fms{n}\gamma_5 
q''_{\n}\big]_u |0 \rangle &=& -2i f_P E_P \phi_P (u,\mu),
\label{phiP}\\   
\langle V_L | \big[\overline{q}'_{\n} \fms{n} 
q''_{\n}\big]_u |0 \rangle &=& 2i f_V E_V \phi_V (u,\mu), \label{phiV} 
\end{eqnarray} 
where $P$ and $V_L$ denote pseudo-scalar and longitudinal
vector mesons respectively. Transversely polarized mesons, $V_T$,
do not contribute at leading order, as in exclusive two-body $B$
decays (for charming penguins, see below). Thus the decay amplitude
can be written as  
\begin{equation} 
\langle X M | H_I | B \rangle 
= i\frac{2G_F}{\sqrt{2}} f_M  E_M
\int^1_0 du ~\T_M^{(q)} (u,\mu) ~\phi_M (u,\mu) 
\langle X | \overline{q}_n \fms{\overline{n}} P_L
  Y_n^{\dagger} b_v | B \rangle, 
\label{amp2}
\end{equation}
where the hard kernels $\T_M^{(q)}$ for various processes are listed in
Table \ref{table:TM}.  
\begin{table}
\begin{tabular}{cccc}\hline\hline
Mode ($\Delta S=1$) & $\T_{M}^{(s)}$ & Mode ($\Delta S=0$) &
$\T_{M}^{(d)}$ \\ \hline 
$ K^{(*)-} X^+$ & $\lambda_p^{(s)} (\C_{1}^{p}+ \C_{4}^{p}) $ 
&$\pi^-X^+/\rho^- X^+$  &  $\lambda_p^{(d)} (\C_{1}^{p}+ \C_{4}^{p}) $  
\\
$ \overline{K}^{(*)0} X^-$  &  $\lambda_p^{(s)} \C_{4}^{p} $ & 
$K^{(*)0}X_s^0/X_s^-$ & $\lambda_p^{(d)} \C_{4}^{p} $  \\ 
$\phi X_s^-/\phi X_{s}^0$&  
$\lambda_p^{(s)} \C_{4}^{p}-\lambda_t^{(s)} (\C_5+ \C_6)$ 
&  $\phi X^0/\phi X^-$&   $-\lambda_t^{(d)} (\C_5+ \C_6)$ \\ \hline
$\pi^0 X_{s\bar s}^0$ &  
$\frac{1}{\sqrt{2}}(\lambda_p^{(s)} \C_{2}^{p} +\lambda_t^{(s)}
\C_3)$  & $\pi^0 X_{\bar s}^0$ &   
$\frac{1}{\sqrt{2}}(\lambda_p^{(d)} \C_{2}^{p} 
+\lambda_t^{(d)} \C_3-\lambda_p^{(d)} \C_{4}^{p})$ \\ 
$\rho^0 X_{s\bar s}^0$ &
$\frac{1}{\sqrt{2}}(\lambda_p^{(s)} \C_{2}^{p} -\lambda_t^{(s)} \C_3)$    
& $\rho^0 X_{\bar s}^0$ &   
$\frac{1}{\sqrt{2}}(\lambda_p^{(d)} \C_{2}^{p} 
-\lambda_t^{(d)} \C_3-\lambda_p^{(d)} \C_{4}^{p} )$ \\ 
$ \w X_{s\bar s}^0$ & 
$\frac{\lambda_p^{(s)}}{\sqrt{2}}\C_{2}^{p}-
\frac{\lambda_t^{(s)}}{\sqrt{2}}(\C_{3} +2\C_5+2\C_6)$ 
& $\w X_{\bar s}^0$  & 
$\frac{ \lambda_p^{(d)}}{\sqrt{2}} (\C_{2}^{p}+ \C_{4}^{p})- \frac{
  \lambda_t^{(d)}}{\sqrt{2}}(\C_3+2\C_5+2\C_6)$ 
\\ 
$K^{(*)-} X_{\bar s}^+$  &  
$\lambda_p^{(s)} (\C_{1}^{p}+ \C_{4}^{p}) $
&$ \pi^{-} X_{\bar s}^+/ \rho^{-}X_{\bar s}^+$  &
$\lambda_p^{(d)}(\C_{1}^{p}+ \C_{4}^{p}) $  \\ \hline\hline 
\end{tabular}
\caption{\baselineskip 3.0ex 
Hard kernels $\T_{M}^{(q)}$ for $\overline B^0/B^-\to X M$ (above
horizontal line) and $\overline B^0_s\to X M$ decays (below horizontal
line), where only the strangeness content of the inclusive jet is
shown. The summation over $p=u,c$ is implied. The NLO Wilson
coefficients $\C_i^p$ are given in Appendix~\ref{appA}.}
\label{table:TM}
\end{table}

In order to obtain the decay rates for  $B\to X M$  
\begin{equation}\label{decayrate}
\frac{d\Gamma}{d E_M} = (2\pi)^2\frac{E_M^3 G_F^2
  f_\pi^2}{m_B}\left|\int_0^1 du\T_M^{(q)} \phi_M\right|^2\sum_X 
|\langle X |\overline{q}_n \fms{\overline{n}} P_L
  Y_n^{\dagger} b_v | B \rangle|^2 \delta^4(p_B - p_X - p_M),
\end{equation}
we use the optical theorem to relate the decay rate to the imaginary 
part of the forward scattering amplitude.
We therefore consider the time-ordered product of the 
heavy-to-light currents 
\begin{equation} 
T(E_M) = \frac{i}{m_B} \int d^4 z ~e^{-ip_M\cdot z} 
\langle B |\mathrm{T} J_H^{\dagger} (z)  J_H (0) | B \rangle, 
\label{tem} 
\end{equation}  
where $J_H (z) = e^{i(\tilde{p} - m_b v )\cdot z}~\overline{q}_n  
\fms{\overline{n}} P_L Y_n^{\dagger} b_v (z)$.
Since there are no collinear particles in the $B$ meson, the
time-ordered product of the collinear fields can be written as 
\begin{equation} 
\langle 0 | \mathrm{T} q_n (z) \cdot \bar{q}_n  (0) | 0
\rangle = i \frac{\fms{n}}{2} \delta({n\cdot z}) \delta^2(z_{\perp}) 
\int \frac{dn\cdot k}{2\pi} e^{-in\cdot k \n\cdot z/2} 
J_P(n\cdot k +i\epsilon),
\label{Jet}
\end{equation} 
which defines the jet function $J_P =J_P (n\cdot k)$ with the label
momentum $P$. In $\SII$, the remaining matrix elements are
parameterized in terms of the $B$ meson shape function,
\beq
\begin{split}
f(n\cdot l ) &= \half \int \frac{d\n\cdot z}{4\pi} e^{-in\cdot l
  \n\cdot z/2} \langle B_v | \Bigl[\overline{b}_v Y_n \Bigr] 
(\n\cdot z/2) \Bigl[ Y_n^{\dagger} b_v \Bigr](0) | B_v \rangle \\ 
&= \half \langle B_v | \overline{b}_v Y_n \delta(n\cdot l
- n\cdot i\partial) Y_n^{\dagger} b_v | B_v \rangle, \label{shape}
\end{split}
\eeq
and the time-ordered product in Eq.~\eqref{tem} can be written as  
\begin{equation} 
T(E_M) = -2 \int^{\overline{\Lambda}}_{-m_b(1-x_M)} dn\cdot l
~f(n\cdot l) ~J_P \Bigl(m_b(1-x_M) + n\cdot l + i\epsilon\Bigr),
\label{tem1}
\end{equation} 
with the limits on $n\cdot l$ included according to
Eq.~\eqref{ndotl}. Taking the discontinuity, we obtain
\begin{eqnarray} \label{Sx} 
\frac{1}{\pi} \mathrm{Im} \, T(E_M) &=& 
2 \int^{\overline{\Lambda}}_{-m_b(1-x_M)} dn\cdot l
~f(n\cdot l) \Bigl[-\frac{1}{\pi}\mathrm{Im}J_P \Bigl(m_b(1-x_M) +
n\cdot l + i\epsilon\Bigr) \Bigr] \\
&\equiv& \frac{2}{m_b} \mS(x_M,\mu_0), \nonumber 
\end{eqnarray}  
where the nonperturbative function $\mS$ is defined as the convolution
of the $B$ meson shape function and the imaginary part of the jet
function. 

Combining Eqs.~\eqref{amp2} and \eqref{Sx}, the factorized
differential decay rate for the $B\to X M$ is 
\begin{equation}  
\frac{d\Gamma}{dE_M}(B\to X M) = \frac{G_F^2}{8\pi} f_M^2 m_b^2 x_M^3 
\mS(x_M, \mu_0) H_M^{(q)}(m_b,\mu_0), 
\label{decay}
\end{equation} 
where $H_M^{(q)}$ is the convolution of the hard kernel and the LCDA,
\begin{equation} 
H_M^{(q)} (m_b, \mu_0) = 
\Big|\int^1_0 du \T_M^{(q)} (u,\mu_0) \phi_M(u,\mu_0)\Big|^2.  
\end{equation}
The information on the LCDA can in principle be extracted from
experimental data on other hard processes, while $H_M^{(q)}$ can be
computed in perturbation theory. It is worth mentioning that
Eq.~\eqref{decay} is independent of $\mu_0$ and $\mu$. $\mS(x_M,
\mu_0)$ is the convolution of the imaginary part of the jet function,
which is computed in matching between $\SI$ and $\SII$ at $\mu_0$ and
evolves down to $\mu$, with the $B$-meson shape function, evaluated at
$\mu$. The dependence on the low scale $\mu$ cancels between the
two. In $H_M^{(q)}$, $\T_M^{(q)}$ evolves from $m_b$ to $\mu_0$,
and the LCDA $\phi_M$, which is the matrix element of the collinear
quark operators, are evaluated at $\mu_0$. The dependence on $\mu_0$
in $H_M^{(q)}$ will then cancel against $\mathrm{Im}J_P$. Therefore
the decay rate is independent of $\mu_0$ and $\mu$.  

We can compare our result with  the differential decay rate for 
$\overline{B} \to X_s \gamma$ in the endpoint region at leading order
\cite{Bauer:2000ew,Bauer:2001yt},
\begin{equation} 
\frac{d\Gamma}{dE_{\g}}(\overline{B} \to X_s \g) = 
\frac{G_F^2 m_b^4}{16\pi^4} x_{\gamma}^3 \alpha  H_{\g} (m_b, \mu_0)
\mS(x_{\g},\mu_0),
\end{equation} 
where $x_{\gamma}=2E_{\gamma}/m_b$, and $\alpha$ is the fine structure
constant. $H_{\g}$ is the hard coefficient
\begin{equation} 
H_{\g} (m_b,\mu_0) = |V_{tb}V^*_{ts}|^2 |C_{\g} |^2 
|C_{1} +C_{2}|^2,
\end{equation}  
with $C_1+ C_2 = 1 + O(\alpha_s)$. Here we have used the operator
basis suggested in Ref.~\cite{Chay:2005ck} for the Wilson
coefficients, which is equivalent to the one in
Ref.~\cite{Bauer:2000yr}. Note that $\mathcal{S}$, the convolution of
the jet function and the $B$ meson shape function, appears exactly as
in $B\rightarrow XM$. Therefore if we take the ratio of these two
decays, this factor cancels out, reducing the theoretical
uncertainty. In the $SU(3)$ limit, the ratio is given by
\begin{equation} 
\Biggl[\frac{d\Gamma(B\to X M)}{dE_M}\Biggl/
\frac{d\Gamma(B\to X_s \g)}{dE_{\g}}\Biggr]_{x_M=x_{\g}}=
~\frac{2\pi^3 f_M^2}{\alpha m_b^2} \frac{H_M^{(q)}(m_b,\mu_0)}
{H_{\g}(m_b,\mu_0)},
\label{ratio}
\end{equation} 
which is only a function of the hard coefficients (with $H_M^{(q)}$
including the convolution with the LCDA).  The ratio does not depend
on detailed information about the $B$-meson shape function.  If
charming penguins are present, this result is modified as discussed in
the next section. 

\section{Charming Penguins}\label{charmpengsec}
The size of the nonperturbative charming penguins in
two-body $B$ decays has been debated recently. Semi-inclusive
decays $B\to XM$ can lead to new insight. Unlike two-body $B$
decays, where additional nonperturbative parameters related to the
$B\to M$ form factors enter the predictions, the only
nonperturbative parameters in Eq.~\eqref{ratio} are the
LCDA. If there are experimental deviations from Eq.~\eqref{ratio} that
exceed the uncertainties from subleading corrections when we compare
processes with and without charming penguins (such as $\overline B
\rightarrow X^0 \phi$), they would then unambiguously
confirm the nonperturbative nature of the charm contribution. 

\begin{figure}[b]
\begin{center}
\vspace{2.0cm}
\epsfig{file=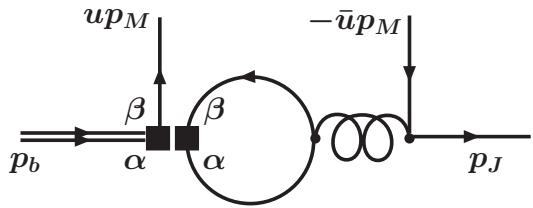, width=7cm}
\end{center}
\vspace{-0.5cm}
\caption{\baselineskip 3.0ex 
A typical charming penguin contribution, with charm quarks in the
loop, and $\alpha$, $\beta$ are the  
color indices. The outgoing momenta $p$ and $p_M$ are $n$ and  
$\bar{n}$-collinear, respectively. Usoft interactions are not shown.} 
\label{charm}
\end{figure}

A typical charming penguin contribution is shown in
Fig.~\ref{charm}. When the momentum transfer through the gluon is
close to $4m_c^2$, the intermediate charm quark pair is almost
on-shell and should be treated  nonperturbatively, governed 
by usoft interactions.  The long-distance contribution can be
power counted as leading order in SCET \cite{Bauer:2005wb} 
and cannot be disentangled from the bound state of the bottom
quark. We can write the momentum of the charm quark 
pair as $2m_c v_{\bar cc}^{\mu} + k^{\mu}$, where  $v_{\bar cc}^{\mu}$
is the velocity of the charm quark pair, and $k^{\mu}$ is the residual
momentum of order $\Lambda_{\mathrm{QCD}}$. Note that 
$v_{\bar  cc}^{\mu}$ is not the usual small velocity parameter $v^*$
in NRQCD. In the rest frame of the $B$ meson, we can write
\begin{equation} 
2m_c v_{\bar cc}^{\mu} \sim \n\cdot p \frac{n^{\mu}}{2} + \bar{u}
n\cdot p_M  \frac{\n^{\mu}}{2},
\label{cvv} 
\end{equation} 
where $v_{\bar cc}^2 =1$. The momentum fraction of $\bar{u}$ of the
antiquark in meson $M$ is given as     
\beq\label{rEq}
4r^2 = \bar{u} x_M, \qquad r\equiv m_c/m_b,
\eeq
where $\bar{u} =1-u$ and $x_M$ is close to 1 near the endpoint.

\begin{figure}[b]
\begin{center}
\vspace{1.4cm}
\epsfig{file=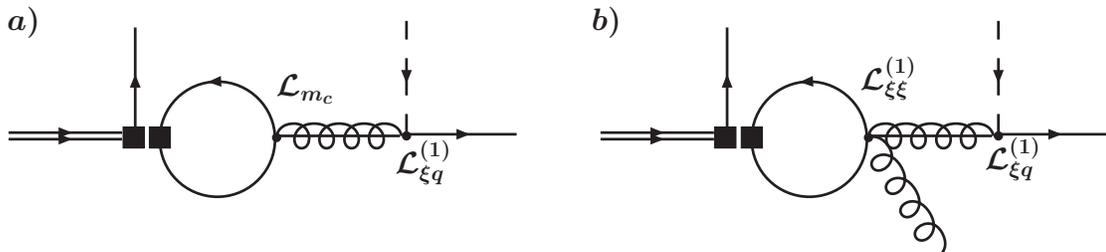, width=14cm}
\end{center}
\vspace{-1.0cm}
\caption{\baselineskip 3.0ex
a) Nonperturbative penguins where charm quarks in the loop are treated as
hard-collinear, with an insertion of $\mathcal{L}_{m_c}$ of order
$\lambda^0$ in $\SI$. b) one of the nonperturbative 
penguin contributions for massless quarks with the insertion of
$\mathcal{L}_{\xi q}^{(1)}$ and $\mathcal{L}_{\xi\xi}^{(1)}$, each of
which is suppressed by $\lambda$ in $\SI$. Dashed line is an usoft
quark, leading to additional endpoint suppression in forming a meson
$M$.  } 
\label{charmlight}
\end{figure}

There are two possible scenarios when we take appropriate
limits of the charm quark mass compared to $m_b$. First, we can take 
the heavy quark limit $m_b \to \infty$ with $m_c \sim
\sqrt{m_b\Lambda}$. In this case, $r^2$ is of order
$\Lambda/m_b$ and the momentum of the charm pair, or the charm quark
itself, becomes hard-collinear in the $n$ direction because $\bar{u} =
4r^2/x_M$ is of order $\Lambda/m_b$. Therefore the outgoing antiquark
with the momentum $\bar{u}p_M$ is  an usoft quark, and the exchanged
gluon has offshellness of order $\sqrt{m_b\Lambda}$. 
The long-distance charming penguin contribution, shown in
Fig.~\ref{charmlight} a), can be treated in
the same way as that of the quarks with small mass, shown in
Fig.~\ref{charmlight} b).  
Since the usoft-collinear interaction is suppressed at least by 
$\lambda \sim \sqrt{\Lambda/m_b}$, the leading long-distance
contribution is suppressed by $m_c/m_b$ at the operator level. In
addition, the unbalanced endpoint configuration for the meson $M$  
gives the endpoint suppression of order $\Lambda/m_b$. Therefore, the
long-distance contribution in this limit is suppressed by
$m_c\Lambda/m_b^2$. This power counting is in agreement with
the expectation that the contribution of the charming penguin 
in this limit gives the same contribution as massless quarks,
which is suppressed by $\Lambda^2/m_b^2$ 
\cite{Feldmann:2004mg}. The reason why it is not of order $m_q
\Lambda/m_b^2$ is because the quark mass insertion is replaced by
the insertion of $\mathcal{L}_{\xi\xi}^{(1)}$.   
        
The second scenario is to take the heavy quark limit $m_b,~m_c \to \infty$ 
with fixed $r$, which is motivated by the fact that $\overline{u} = 
4r^2/x_M$ is near the central point experimentally. In this case, the
power counting is different. The charm quark is regarded as a heavy
quark and there is no endpoint suppression. The exchanged 
gluon in Fig.~\ref{charm} has large offshellness $4m_c^2 \sim
m_b^2$, and is integrated out to obtain an operator of the form 
$[\bar{c}c]_{\mathrm{NR}}[\bar{\xi}_{\n}\xi_{n}]$. Here we suppress the
Dirac structure and color indices, and the charm
quark is treated as a heavy quark described by NRQCD. The
nonperturbative charming penguin contribution is then 
obtained from the time-ordered product of this operator with the 
four-quark operators in the weak Hamiltonian  of the form $[\bar{\xi}_n
b_v][\bar{c}{c}]_\mathrm{NR}$. This leads to a contribution $\sim
\alpha_s (2m_c) v^* \mathcal{F}_{c\bc}$ in agreement with   
Refs.~\cite{Bauer:2005wb,Bauer:2004tj}, 
where the nonperturbative function $\mathcal{F}_{c\bc}$ 
is comparable in size to the leading-order shape 
function $f$. Furthermore there is no endpoint suppression because the
outgoing antiquark forming the meson $M$ is collinear with 
momentum fraction $\bar{u} = 4r^2/x_M $, corresponding numerically to the 
the central region in the LCDA of the meson $M$.

We think that the second scenario is more plausible based on the
actual value of $m_c/m_b$. As this is the more conservative of the
two, we consider the nonperturbative charming penguins in the second
scenario, which give larger contributions than the first scenario. As
explained above, in the heavy quark limit $m_{b,c} \to \infty$ with
$r$ fixed, the charming penguin can be of leading order, which   
can be expanded in powers of $\Lambda/m_{b,c}$ and $\alpha_s(2m_c)$
in a consistent way.  In this scheme, at leading order in
$1/m_{b,c}$, the contribution is factorized into the
$\n$-collinear part, the jet function, and a nonperturbative function.  
The derivation of the factorized form is presented in detail in
Appendix~\ref{appB}. At leading order in $\Lambda/m_{b,c}$, and to
first order in $\alpha_s (2m_c)$, the contribution of the
charming penguin to the differential decay rate can be written as [see
Eq.~(\ref{facc})]  
\begin{equation} 
\frac{d\Gamma_{\bar{c}c}(B\to X_q M)}{dE_M} 
= \frac{G_F^2}{8\pi} f_M^2 m_b^2 x_M^3 
\lambda_c^{(q)}  \alpha_s(2m_c)c_q^{BM}
\phi_M \Bigr(u=1-\frac{4r^2}{x_M}\Bigr) 
\cdot 2\mathrm{Re} \T_M^{(q)*} {\cal F}_{c\bc}, 
\label{dcharm}
\end{equation}   
where $c_{q}^{BM}$ is defined in Appendix \ref{appB} and $\T_M^{(q)}$
is the hard kernel given in Table \ref{table:TM}.
Here $M=P,V_L$, while the contribution of $V_T$ is $1/m_{b,c}$
suppressed because of the spin flip. The function $\mathcal{F}_{c\bc}$
does not depend on the outgoing meson $M$ or the jet $X$ because
$\mathcal{F}_{c\bc}$ is given by the nonperturbative 
effects arising only from the usoft interactions of the on-shell
charm quark pair in the $B$ meson. 
Hence, up to the $B$ meson flavor,  $\mathcal{F}_{c\bc}$ is
universal in all the decay modes where charm penguins contribute, and
its size is experimentally measurable from various decay modes.
In the isospin limit the $\mathcal{F}_{c\bc}$ functions in $\overline
B^0$ and $B^-$ decays are the same, and are equal to
$\mathcal{F}_{c\bc}$ in $B_s$ decays in the $SU(3)$ limit. Due to its
nonperturbative nature it can, however, have a nonzero strong phase
\cite{Bauer:2004tj}. 

In summary, the differential decay rate for the semi-inclusive decays 
at leading order in $1/m_{b}$, including the nonperturbative charming
penguin at LO in $1/m_c$ and $\alpha_s(2m_c)$, is
\begin{eqnarray} \label{decaylo}
\frac{d\Gamma}{dE_M} (B\to X_q M) &=& \frac{G_F^2}{8\pi} f_M^2 m_b^2
x_M^3 \Biggl[H_M^{(q)}(m_b,\mu_0)  \mS(x_M, \mu_0) \\ 
&&
+\alpha_s(2m_c) \lambda_c^{(q)}c_{q}^{BM}
\phi_M\left(1-\frac{4r^2}{x_M}\right) 
2\mathrm{Re} \T_M^{(q)*} \mathcal{F}_{c\bc} \Biggr]. \nonumber
\end{eqnarray}
Phenomenological implications of this expression will be discussed in
section \ref{pheno}.

\section{Estimates of subleading corrections}\label{beyondLO}
To predict the branching ratios for $B\to XM$ more accurately,
Eq.~\eqref{decaylo} can be systematically extended to higher orders in
$1/m_b$ and $\alpha_s$. In this way one could also unambiguously
determine whether a possible future discrepancy between experiment and
predictions based on Eq.~\eqref{ratio} is due to nonperturbative
charming penguins or higher-order corrections. A full analysis of the
higher-order corrections is beyond the scope of this paper, but we
identify typical subleading corrections and estimate their size.
The subleading corrections, suppressed by powers of $\Lambda/m_b$, are
of two types: (i) corrections to the heavy-to-light current 
leading to  the subleading $B$-meson shape 
functions, some of which are already well known from the analyses  of
$\overline{B}\to X_s\g$ and $\overline{B}\to X_u l\bar{\nu}$ inclusive 
decays
\cite{Bauer:2001mh,Leibovich:2002ys,Bauer:2002yu,Lee:2004ja,Bosch:2004cb},   
and (ii) corrections to the $\n$-collinear currents forming
the light meson $M$, which appear as the twist-3 LCDA and the $SU(3)$
breaking effects in the twist-2 LCDA. 

Let us consider the corrections of the first type. The
convolution $\mathcal{S}$ of the jet function  and the $B$ meson shape 
function in Eq.~\eqref{Sx} can be expanded to higher orders in $1/m_b$
as  
\begin{equation} 
\mS(x_M) = \mS^{(0)} (x_M) + \Bigl( \mS^{(1)}_{hl} (x_M) 
+ \mS^{(1)}_{\n} (x_M) \Bigr) +\cdots,
\end{equation}  
where $\mS_{hl}^{(1)}$ is the subleading corrections to the
heavy-to-light current and $\mS_{\n}^{(1)}$ is the usoft corrections
to the $\n$-collinear current. Using the results of
Ref.~\cite{Bosch:2004cb}, the sum 
$\mS^{(0)}+\mS_{hl}^{(1)}$ can be related to the imaginary part of the
time-ordered product, $W_{\mu\nu}$, 
\begin{equation} 
\frac{1}{m_b} \Bigl(\mS^{(0)}+\mS_{hl}^{(1)}\Bigr) = W_{\mu\nu} 
\frac {\n^{\mu}\n^{\nu}}{4},
\end{equation} 
where the factor $\n^{\mu}/2$ comes from the
$\n$-collinear current $\langle \bar{\xi}_{\n} \g^{\mu} \xi_{\n}
\rangle =\n^{\mu} \langle \bar{\xi}_{\n} \fms{n} \xi_{\n} \rangle /2$. 
Taking the ratio with respect to  $\overline{B}\to X_s
\gamma$ gives the difference
\begin{eqnarray} 
\mS_{hl}^{(1)} - \mS_{\g}^{(1)} &=& \Bigl(\mS^{(0)} +\mS_{hl}^{(1)}
\Bigr) -  \Bigl(\mS^{(0)} +\mS_{\gamma}^{(1)} \Bigr) \\
&\sim& 2\int dn\cdot l ~v(n\cdot l) ~\delta(m_b (1- x)+n\cdot l)
= -\frac{2}{m_b^2} H_2(1- x), \nonumber 
\end{eqnarray} 
where $x=x_M=x_{\g}$ is  chosen. The subleading shape functions $v$
and $H_2$ are defined in Refs.~\cite{Bosch:2004cb} and
\cite{Bauer:2001mh}, respectively. In particular, $H_2$ is
proportional to $\lambda_2 = \langle \bar{B}_v | \bar{b}_v g_s \s
G^{\mu\nu} b_v | \bar{B}_v \rangle/12 \sim 0.12 \ \mathrm{GeV}^2.$
Taking a broad cut  $E_M \ge  2.0$ GeV, this contribution should
not exceed $10\%$ compared to the leading contribution, unless there
is an enhancement in the coefficient.  

The subleading correction $\mS_{\n}^{(1)}$ comes from the usoft
interactions with the $\n$-collinear currents, which lead to new
subleading $B$ meson shape functions. The subleading operators
obtained by inserting the usoft gauge-invariant term
$Y_{\n}^{\dagger}i\fmsl{D}^{\perp}_{us}  Y_{\n}$ are
suppressed by  $\lambda^2$ in $\SI$, but they should be included in
$\SII$ because they are suppressed by $\Lambda/m_b$. The nonzero
contributions come only from the color-octet four-quark operators   
\beq 
\begin{split} \label{subleadingObar}
{\cal O}_{iA}^{(1)}(u) &=2
\big(\bar{q}_{n} Y_n^{\dagger} Y_{\n} \gamma^\mu P_L T^a Y_{\n}^{\dagger}
b_v \big)\cdot 
\Big[\bar{q'}_{\n} \,\gamma_\mu\, P_{L,R}\, T^a
\,Y_{\n}^{\dagger}i\fmsl{D}^{\perp}_{us}  Y_{\n}  
\,\frac{1}{n\cdot {\cal P}}\,\fms{n} \,q''_{\n}\Big]_u,
\\
{\cal O}_{iB}^{(1)}(u) &=2
\big(\bar{q}_{n} Y_n^{\dagger} Y_{\n} \gamma^\mu P_L T^a Y_{\n}^{\dagger}
b_v \big)\cdot\Big[\bar{q'}_{\n} \;\fms{n} \frac{1}{n\cdot {\cal
    P}^\dagger} Y_{\n}^{\dagger}i\overleftarrow{\fmsl{D}}^{\perp}_{us}
Y_{\n}\, T^a \gamma_\mu\, P_{L,R} \, q''_{\n}\Big]_u,
\end{split}
\eeq
where the flavor and chirality structure is the same as the
corresponding leading operators ${\cal   O}_i$ in Eq.~\eqref{siop}.  
Because of reparameterization invariance
\cite{Chay:2002vy,Manohar:2002fd}, the Wilson coefficients  of
${\cal O}_{i,A}^{(1)}(u) +{\cal O}_{i,B}^{(1)}(u)$ are the same as
those for the leading color-octet operators in Eq.~\eqref{octet},
which are presented in Appendix~\ref{appA}.

After some calculation, the matrix elements of these operators
which do not vanish trivially from the flavor content are
nonzero only for specific values of $i$ due to the helicity structure. 
They are given by 
\begin{eqnarray} 
\label{oa82}
\langle {\cal O}_{i,A}^{(1)}(u) \rangle &=& \frac{if_M}{N} \frac{\phi_M
  (u)}{\overline{u}} \langle X_q | \overline{q}_n  \Bigl[Y_n^{\dagger}
i\fmsl{D}^{\perp}_{us}   
Y_{\n}\Bigr] P_L Y_{\n}^{\dagger} b_v | B \rangle, \ (i=1,2,4,5),
\\ 
\label{ob82}
\langle {\cal O}_{i,B}^{(1)}(u) \rangle &=& (\delta_{VM}-\delta_{PM})
\frac{if_M}{N} \frac{\phi_M (u)}{u}
\langle X_q | \overline{q}_n Y_n^{\dagger} Y_{\n}
\Bigl[Y_{\n}^{\dagger}i\overleftarrow{\fmsl{D}}^{\perp}_{us}\Bigr] 
P_L b_v | B \rangle, \ (i=3,6),  \nonumber   
\end{eqnarray}
where  $\delta_{PM}$, $\delta_{VM}$ are Kronecker deltas, and we
use Eqs.~\eqref{wave} and \eqref{vtom} in Appendix~\ref{appB} for the
matrix elements of the collinear current. For the matrix elements of the
heavy-to-light current, we need the time-ordered products 
with $J_H^{\dagger}$, 
\begin{eqnarray}
T_A^{(2)} &=& \frac{2i}{m_b m_B} \int d^4 z ~e^{-ip_M\cdot z} 
\langle B |\mathrm{T} J_H^{\dagger} (z) ~ 
\overline{q}_n \Bigl[Y_n^{\dagger} i\fmsl{D}^{\perp}_{us}  
Y_{\n}\Bigr] P_L Y_{\n}^{\dagger} b_v (0) | B \rangle,
\\   
T_B^{(2)} &=& \frac{2i}{m_b m_B} \int d^4 z ~e^{-ip_M\cdot z} 
\langle B |\mathrm{T} J_H^{\dagger} (z) ~ 
\overline{q}_n Y_n^{\dagger} Y_{\n}
\Bigl[Y_{\n}^{\dagger}i\overleftarrow{\fmsl{D}}^{\perp}_{us}\Bigr] 
P_L b_v (0) | B \rangle. \nonumber    
\end{eqnarray}
They can be factorized into the jet function and subleading
$B$-meson shape functions,
\begin{equation} 
T_{A,B}^{(2)} =  -2 \int^{\overline{\Lambda}}_{-m_b(1-x_M)} dn\cdot l 
~f^{(1)}_{A,B} (n\cdot l) ~J_P \Bigl(m_b(1-x_M) + n\cdot l +
i\epsilon\Bigr), 
\end{equation} 
where the subleading shape functions are defined as 
\begin{eqnarray}\label{fA1}
f_A^{(1)} (n\cdot l) &=& \frac{1}{m_b} 
 \langle B_v | \bar{b}_v Y_n \delta(n\cdot l - n\cdot
 i\partial) \frac{\fms{\n}\fms{n}}{4}
\Bigl[Y_n^{\dagger} i\fmsl{D}^{\perp}_{us}  
Y_{\n}\Bigr] Y_{\n}^{\dagger} b_v | B_v \rangle,
\\   
f_B^{(1)} (n\cdot l) &=& \frac{1}{m_b} 
 \langle B_v | \bar{b}_v Y_n \delta(n\cdot l - n\cdot
 i\partial) \frac{\fms{\n}\fms{n}}{4} Y_n^{\dagger} Y_{\n}
\Bigl[Y_{\n}^{\dagger}i\overleftarrow{\fmsl{D}}^{\perp}_{us}\Bigr] 
 b_v (0) | B_v \rangle.   \nonumber  
\end{eqnarray}
The shape functions $f_{A,B}^{(1)}$ are different from the subleading
shape functions appearing in $\overline{B}\to X_s \g$ and
$\overline{B}\to X_u l\bar{\nu}$, due to the presence of
$Y_{\n}^{(\dagger)}$, which cannot be neglected at subleading
order. At present we cannot estimate their size, but there is no
reason to expect a dramatic enhancement from the insertion of 
$Y_{\n}^{(\dagger)}$. However, these contributions can be numerically
significant in the decays that are very small at LO in $1/m_b$. These
are the color-suppressed $\Delta S=0$ tree decays $\overline
B_s^0\to\{\pi^0,\rho^0,\omega\} X_{\bar s}^0$, the QCD
penguin-dominated $\Delta S=0$ and $\Delta S=1$ 
decays $\overline{B}^0\to \phi X^0, B^-\to \phi X^-$ and $\overline
B_s^0\to\omega X_{s \bar s}^0$ and the $\lambda_u^{(s)}$ part of
the amplitudes in $\overline B_s^0\to\{\pi^0,\rho^0\} X_{s \bar s}^0$. 
For these decays, the Wilson coefficients of the LO operators
convoluted with the asymptotic LCDA ($|\phi\otimes\C_2^u|\sim 0.08$ and
$|\phi\otimes(\C_5+\C_6)| \sim 0.005$) are much smaller than those for
the color-octet operators ($|\phi\otimes\C_2^u|\sim 1.9$ and
$|\phi\otimes(\bar\C_5+\bar\C_6)| \sim 0.15$). The subleading
contributions can thus be numerically large in spite of the $1/2N$
suppression. Note that this is not a sign of failure of the $1/m_b$
expansion, but due to the hierarchy of the Wilson coefficients. 

\begin{figure}[t]
\begin{center}
\epsfig{file=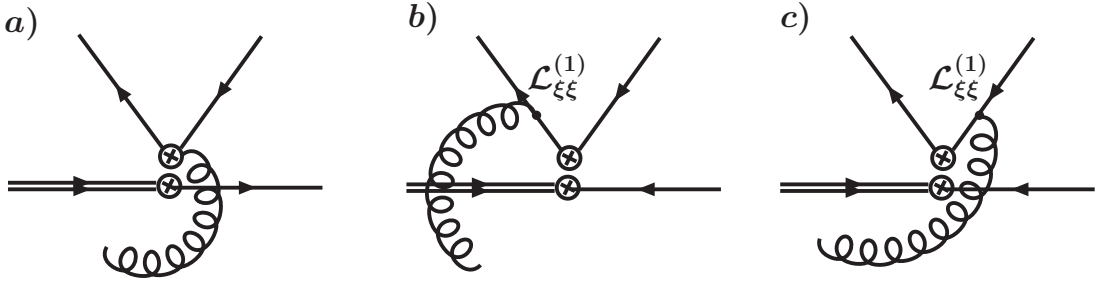, width=12cm}
\end{center}
\vspace{-1.0cm}
\caption{\baselineskip 3.0ex 
Subleading usoft interactions induced from the $\n$-collinear currents. 
In Diagrams b) and c), the dots denote the $\lambda$-suppressed
interaction in $\mathcal{L}_{\xi\xi}^{(1)}$. }
\label{usoft}
\end{figure}

There is another contribution  shown in Fig.~\ref{usoft},
from the time-ordered products of the $\n$-collinear currents
and $\mathcal{L}_{\xi\xi}^{(1)}$, 
\begin{equation} 
\mathcal{L}_{\xi\xi}^{(1)} = \bar{\xi}_{\n} W_{\n} 
\Bigl(Y_{\n}^{\dagger} i\fmsl{D}_{us}^{\perp} Y_{\n} \Bigr) 
\frac{1}{n\cdot \mP} W_{\n}^{\dagger} i\fmsl{D}_{\n}^{\perp}\nn \xi_{\n} 
\label{lxi1}
+\bar{\xi}_{\n} i\fmsl{D}_{\n}^{\perp}\nn W_{\n} \frac{1}{n\cdot \mP} 
\Bigl(Y_{\n}^{\dagger} i\fmsl{D}_{us}^{\perp} Y_{\n} \Bigr) 
\nn W_{\n}^{\dagger}\xi_{\n}.
\end{equation}  
The intermediate legs in Fig.~\ref{usoft} are hard-collinear and give 
a jet function of the form $1/\n\cdot k$. 
The relevant LCDA for $M$ are suppressed by $\Lambda/m_b$ due to the
presence of $iD_{\n}^{\perp}$. But it is not known whether this
process factorizes, and we leave a full analysis for future work.

We now consider the contributions from the $(S-P)\times (S+P)$
four-quark operator. At leading order it matches onto $\overline{q'}_n
(1-\g_5) b\cdot \overline{q}_{\n}(1+\g_5) 
q_{\n}'$, which vanishes because of spin symmetry. At subleading order
it matches onto  
\beq
\begin{split}
{\cal O}^{(1)}_{S+P} 
= -2 \sum_{q'=u,d,s}&
\big(\overline{q'}_n \, \fms{\n} \,P_R Y_n^{\dagger} b_v\big)
\Big( \overline{q}_{\n} \, P_R 
\frac{1}{n\cdot \mP} W_{\n}^{\dagger} i\fmsl{D}_{\n}^{\perp} W_{\n}
\,\fms n\, {q'}_{\bar{n}}\Big) + \\
+&\big(\overline{q'}_n \, \fms{\n} \,P_R Y_n^{\dagger} b_v\big)
\Big( \overline{q}_{\n} \,\fms n\,\, W_{\n}^{\dagger}
i\overleftarrow{\fmsl{D}}_{\n}^{\perp} W_{\n} \frac{1}{n\cdot
  \mP^{\dagger}}P_R  {q'}_{\bar{n}}\Big), 
\end{split}
\eeq
where for the heavy-to-light current we have used the relation
($2v^{\mu} = n^{\mu}+\n^{\mu}$) 
\begin{equation}  
2\overline{q}'_n P_L Y_n^{\dagger} b_v 
= 2\overline{q}'_n P_L \fms{v} Y_n^{\dagger} b_v
= \overline{q}'_n  \fms{\n} P_R  Y_n^{\dagger} b_v.
\end{equation} 
In semi-inclusive $B$ decays, the amplitude from this operator
factorizes using the twist-3 LCDA \cite{Hardmeier:2003ig}. 
At order $1/m_b$ it contributes through the time-ordered product with
the leading heavy-to-light current as
\begin{equation} 
T_{S+P} = \frac{i}{m_B} \int d^4 z ~e^{-ip_M\cdot z} 
\langle B |\mathrm{T} J_H^{\dagger} (z) ~\overline{q}_n' \fms{\n} P_R
Y_{\n}^{\dagger} b_v (0) | B \rangle.     
\end{equation} 
This vanishes because
\begin{equation} 
\langle B|\mathrm{T} J_H^{\dagger} (z) ~ 
\overline{q}_n \fms{\n} P_R Y_{\n}^{\dagger} b_v (0)| B \rangle  
\propto \langle B|
 \overline{b}_v Y_n (\tfrac{\n \cdot z}{2}) \fms{\n} P_L
 \,\fms{n}\,\fms{\n}P_R   Y_{\n}^{\dagger} b_v (0) |B\rangle= 0.    
\end{equation} 
Therefore the nonzero contribution comes 
from the time-ordered products of $\mathcal{O}_{S+P}^{(1)}$ with the
subleading operators of $J_H^{\dagger}$, suppressed by  $m_q/m_b$
\cite{Chay:2005ck}, or from the time-ordered products of
$\mathcal{O}_{S+P}^{(1)}$ with itself. Both are of order $1/m_b^2$,
but the latter may not be numerically negligible. In the QCD
factorization approach \cite{Beneke:1999br}, the contributions
corresponding to ${\cal O}^{(1)}_{S+P} $ lead to formally suppressed
but numerically large ``chirally enhanced'' contributions. In SCET
${\cal   O}^{(1)}_{S+P} $ is also formally suppressed, while its matrix 
elements remain unknown and could be numerically large. Because of
this uncertainty, the decay rates and the CP asymmetries presented in
the next section for modes with small tree-level amplitudes should be
regarded only as a rough estimate. 

\begin{figure}[b]
\begin{center}
\epsfig{file=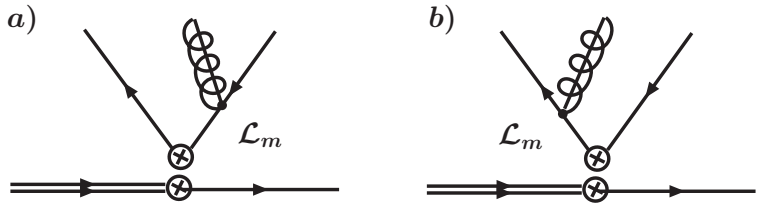, width=10cm}
\end{center}
\vspace{-1.0cm}
\caption{\baselineskip 3.0ex 
The insertion of the strange quark mass in the $\n$-collinear
currents in $\SII$. Diagrams a) and b) represent the possible
strange quark mass insertions for the strange quark in
Eq.~(\ref{su3}) where the external particle is a strange quark.}  
\label{mass}
\end{figure}

Finally, we discuss the $SU(3)$ breaking corrections to
Eq.~\eqref{ratio}. The $SU(3)$ breaking due to different light-quark
flavors in the inclusive jet is suppressed by $m_s^2/m_b\Lambda$
\cite{Chay:2005ck} and thus negligible, but the $SU(3)$ corrections due 
to the strangeness content of meson $M$ are not negligible. These are
realized in SCET by inserting the strange quark mass term
\cite{Leibovich:2003jd,Rothstein:2003wh} 
\begin{equation} 
\mathcal{L}_m = m_s~ \overline{\xi}_{\bar{n}}
\Biggl[i\fmsl{D}_{\n}^{\perp}, \frac{1}{n\cdot iD_{\n}} \Biggr] \nnn
\xi_{\bar{n}} 
\end{equation}
in the leading $\n$-collinear currents with the final-state $s$ quark in
Fig.~\ref{mass}. It can be written as
\begin{equation} \label{su3}
A_{s}(u) = \frac{-i}{ f_M E_M} \langle M |~\mathrm{T}~ 
\big[\overline{s}_{\bar{n}}  \fms{n}P_L
q_{\bar{n}}(0)\big]_u  
\cdot i\int d^4 z \mathcal{L}_m (z) ~| 0 \rangle. 
\end{equation}
For the final-state $\overline{s}$ quark, a hermitian conjugate of
Eq.~(\ref{su3}), $A_{\bar s}(u)$, is needed with $A_{\bar s}(u) =
A_s(\bar u)$. If $m_s$ is comparable to $iD_{\n}^{\perp}\sim \Lambda$, 
$\mathcal{L}_m$ is of leading order in $\SII$,
suppressed only by $m_s/\Lambda$. 

The $SU(3)$ breaking affects meson decay constants and the LCDAs. One
finds to leading order in $SU(3)$-breaking
$\phi_{\bar{K}}(u)=\phi(u)+A_s(u)$, $\phi_K(u)=\phi(u)+A_s(\bar u)$
and $\phi_\eta(u)=\phi(u)+2\big(A_s(u)+A_s(\bar u)\big)/3 $ for the
LCDA of $K^-(\overline K^0), K^+(K^0)$ and $\eta$, where  $\phi(u)$ is
the pion LCDA. Since $\int_0^1\phi_M(u)du=1=\int_0^1\phi(u)du$ one has a
constraint  
\beq 
\int_0^1 A_s(u)du=0. 
\eeq
It is also straightforward to check that these LCDA satisfy the
relation \cite{Chen:2003fp} $\phi_{\pi}(u,\mu)+ 3 \phi_{\eta} (u,\mu) 
= 2\Bigl[\phi_{K^+}(u,\mu)+  \phi_{K^-} (u,\mu)\Bigl] 
= 2\Bigl[\phi_{K^0}(u,\mu)+  \phi_{\bar{K}^0} (u,\mu)\Bigl].$ 
From power counting one expects the relative size of the $SU(3)$
breaking contribution to be of order   
$m_s/\Lambda \sim 20$\%. Recent 
QCD sum rules predictions can be found in
Refs.~\cite{Khodjamirian:2003xk,Ball:2006wn,Ball:2004rg,Ball:2004ye}.

\section{Phenomenology}\label{pheno}
In this section we collect the predictions for $B\to XM$ decay rates
and direct CP asymmetries, defined as
\beq\label{ACP}
A_{CP}=\frac{{d\Gamma(\overline{B}\to XM)}/{dE_M} - d\Gamma(B\to
  XM)/dE_M}{d\Gamma(\overline{B}\to XM)/dE_M + d\Gamma(B\to XM)/dE_M} ,
\eeq
while treating the charming penguins as perturbative. Once the
experimental data become available, one of the modes can be used 
to determine the nonperturbative charm contribution
$\mathcal{F}_{c\bc}$ and then modify the predictions according to
Eq.~\eqref{decaylo}. To reduce the hadronic uncertainty,
we normalize the branching ratios of $\overline
B^0\to X M $, $B^-\to X M $, and $\overline B_s^0\to X M $ to the
decay rates $\overline B^0\to X_s^0\gamma$, $B^-\to X_s^-\gamma$, 
and $\overline B_s^0\to X_{s\bar s}^0\gamma$ in the endpoint region,
respectively [See Eq.~\eqref{ratio}.] The only remaining 
nonperturbative input is then the light meson LCDAs. Expanding in
terms of the Gegenbauer polynomials,  
\beq 
\phi_M(u, \mu)=6 u\bar u\big[1+\sum_{n=1}^\infty a_n^M(\mu)
C_n^{3/2}(2u-1)\big],
\eeq
we truncate the series at $n=2$ and use isospin symmetry to set
$a_1^M=0$ for mesons not containing a strange quark. We 
fix the remaining coefficients using the results from  QCD sum rules, 
while conservatively doubling the errors quoted in the literature. This
gives at $\mu=2$ GeV:  $a_1^K=0.05\pm0.05$, $a_2^K=0.23\pm0.23$
\cite{Ball:2006wn}, $a_1^{K^*}=0.08\pm0.13$, \cite{Ball:2004rg},
$a_2^\pi=0.09\pm0.15$ \cite{Ball:2004ye}, $a_2^{K^*}=0.07\pm0.08$,
$a_2^\rho=0.14\pm0.15$, $a_2^\phi=0.\pm0.15$ \cite{Ball:2004rg}, while
for the $\omega$ LCDA $a_2^\omega=0.\pm0.2$ is used for lack of
better information.

\begin{table}[b]
\begin{tabular}{lccc}\hline\hline
Mode & Br(Mode)/Br($B\to X_s\gamma$)& ~~Exp. (2-body)~~ & $A_{CP}$
\\\hline\hline 
$\overline B^0\to K^{-} X^+ $   & $0.16\pm0.09\pm0.05$ & $>0.078$ &
$0.30\pm0.16\pm0.01$ \\ 
$\overline B^0\to K^{*-} X^+ $  & $0.28\pm0.16\pm0.07$ & $>0.026$&
$0.31\pm0.16\pm0.02$ \\ 
$\overline B^0\to \phi X_{s}^0 $  & $0.22\pm0.13\pm0.03$ & $>0.034$&
$0.0089\pm0.0050\pm0.0016$ \\ 
\hline 
$B^-\to \overline K^{0} X^- $    & $0.20\pm0.11\pm0.06 $& $>0.067$ &
$0.0097\pm0.0048\pm0.0006$ \\ 
$B^-\to \overline K^{*0} X^- $  & $0.34\pm0.19\pm0.08$ & $>0.040$ &
$0.0084\pm0.0046\pm0.0019$ \\ 
$B^-\to \phi X_{s}^- $  & $0.22\pm0.13\pm0.03$ & $>0.035$ &
$0.0089\pm0.0050\pm0.0016$ \\ 
\hline
$\overline B_s^0\to \pi^{0} X_{s \bar s}^0 $   & $(1.0\pm
0.6\pm0.2)\times 10^{-2}$ & $-$& $-$ 
\\ 
$\overline B_s^0\to K^- X_{\bar s}^+ $   & $0.16\pm0.09\pm0.05$ & $-$
& $0.30\pm0.16\pm0.01$ \\ 
$\overline B_s^0\to \rho^0 X_{s\bar s}^0 $   &
$(2.4\pm1.4\pm0.5)\times 10^{-2}$ & $-$& $-$ 

\\ 
$\overline B_s^0\to \omega X_{s \bar s}^0 $   &
$(2.8\pm3.4\pm0.7)\times 10^{-3}$ & $-$& $-$ 
\\ 
$\overline B_s^0\to K^{*-} X_{\bar s}^+ $ & $0.28\pm0.16\pm0.07$& $-$
& $0.32\pm0.16\pm0.02$ \\ 
\hline  \hline    
\end{tabular}
\caption{\baselineskip 3.0ex 
Predictions for decay rates and direct CP asymmetries for
 $\Delta S=1$ semi-inclusive hadronic decays are given in the second
 and fourth column, respectively. The first errors  
are an estimate of the $1/m_b$ corrections, while the second errors
are due to errors on the Gegenbauer coefficients in the expansion of
the LCDA. The third column gives lower bounds on inclusive decay rates
obtained by summing over measured two-body decays and normalizing 
to $b\to s\gamma$ decay with $E_{\rm min}=2.0$ GeV ($90\% $ CL lower
bounds are used).} 
\label{table:DeltaS=1}
\end{table}

Direct CP asymmetries in Eq.~\eqref{ACP} are nonzero only in the
presence of nonzero strong phases. These are generated by integrating
out on-shell light quarks in a loop when matching full QCD to $\SI$ at
NLO in $\alpha_s$. We therefore use the NLO matching expressions for the
Wilson coefficients $\mathcal{C}_i^p$ at $\mu=m_b$ even though the
evolution to the hard-collinear scale $\mu_0\sim \sqrt{\Lambda m_b}$
is performed at NLL. Note that this running cancels to a large extent in
the ratios of decay rates (only the running of 
$a_n^M(\mu), n\geq 1$ remains), giving in effect
the Wilson coefficients with NLO accuracy at the hard-collinear scale
$\mu_0$.  

For definitiveness, we choose $\mu_0=2$ GeV for the hard-collinear scale,
which corresponds to the experimental cut  $p_X^2<(2{\rm\ GeV})^2$ on 
the inclusive jet invariant mass. The corresponding predictions are
listed in Tables \ref{table:DeltaS=1} and 
\ref{table:DeltaS=0} for $\Delta S=1$ and $\Delta S=0$ decays
respectively. The predicted partial decay widths $d\Gamma(B \rightarrow
XM)/dE_M$ in principle depend on the light meson energy $E_M$. In the
endpoint region the dependence on  
$x_M=2E_M/m_B=1+{(m_M^2-p_X^2)}/{m_B^2} $ is, however, 
a subleading effect,\footnote{Numerically, for
$p_X^2<(2{\rm\ GeV})^2$ one has $E_\pi>2.26$ GeV compared to
$m_{B^0}/2=2.64$ GeV. The same $p_X^2$ cut corresponds to higher $E_M$
cut for heavier mesons, for instance, for $\phi$ mesons the same cut on 
$p_X^2$ corresponds to $E_\phi>2.36$ GeV. Thus $m_B/2-E_M\sim \Lambda$
with $1-x_M\sim O(\Lambda/m_B)$.} so we set $x_M=1$ in Tables
\ref{table:DeltaS=1} and 
\ref{table:DeltaS=0}.  

\begin{table}
\begin{tabular}{lccc}\hline\hline
Mode & Br(Mode)/Br($B\to X_s\gamma$)&  ~~Exp. (2-body)~~
&$A_{CP}$\\\hline\hline 
$\overline B^0\to \pi^- X^+$  &  $0.67\pm0.37\pm0.14 $ & $>0.038$ &
$-0.040\pm0.021\pm0.004 $ \\ 
$\overline B^0\to K^0 X_s^0$  &  $(9.1\pm 5.3\pm 3.1)\times 10^{-3}$
&$>2.0 \times 10^{-3}$ 
&  $-0.15\pm0.11\pm0.01 $ \\ 
$\overline B^0\to \phi X^0$  &  $(2.0\pm2.0\pm0.1)\times 10^{-4} $ &
$-$ &  $-$ \\ 
$\overline B^0\to \rho^- X^+$  &  $1.76\pm0.97\pm0.35 $ & $>0.10$ &
$-0.039\pm0.021\pm0.004$ \\ 
$\overline B^0\to K^{*0} X_s^0$  &  $(1.4\pm0.8\pm0.5)\times
10^{-2} $ &  $-$ &$-0.17\pm0.11\pm0.03$ \\ 
\hline  
$B^-\to K^0 X_s^-$  &  $(9.1\pm 5.3\pm 3.1)\times 10^{-3}$ &$>2.5
\times 10^{-3}$& 
$-0.15\pm0.11\pm0.01 $ \\ 
$B^-\to \phi X^-$  &  $(2.0\pm2.0\pm0.1)\times 10^{-4} $ &  $-$ &  $-$\\
$B^-\to K^{*0} X_s^-$  &  $(1.4\pm0.8\pm0.5)\times 10^{-2} $ &  $-$ &
$-0.17\pm0.11\pm0.03$ \\ 
\hline   
$\overline B_s^0\to \pi^0 X_{\bar s}^0$  &
$(4.1\pm4.1\pm2.6)\times 10^{-3}$ &  $-$&  $-$ 
\\ 
$\overline B_s^0\to \pi^- X_{\bar s}^+$  & $0.67\pm0.37\pm0.14
$ &  $-$ &  $-0.040\pm0.021\pm0.004 $ \\ 
$\overline B_s^0\to \rho^0 X_{\bar s}^0$  &
$(1.3\pm1.3\pm0.7)\times 10^{-2} $ &  $-$&  $-$
\\ 
$\overline B_s^0\to \rho^- X_{\bar s}^+$  & $1.76\pm0.97\pm0.35
$ &  $-$&  $-0.039\pm0.021\pm0.004 $ \\ 
$\overline B_s^0\to \omega X_{\bar s}^0$  &
$(1.1\pm1.1\pm0.9)\times 10^{-2} $ &  $-$&  $-$
\\ 
\hline\hline
\end{tabular}
\caption{\baselineskip 3.0ex 
Predictions for decay widths and direct CP asymmetries of
  $\Delta S=0$ semi-inclusive hadronic decays.   The first errors 
are an estimate of the $1/m_b$ corrections, while the second errors
are due to errors on the Gegenbauer coefficients in the expansion of
the LCDAs. The third column gives lower bounds on inclusive decay
rates obtained by summing over measured two-body decays and
normalizing to $b\to s\gamma$ decay with $E_{\rm min}=2.0$ GeV ($90\%
$ CL lower bounds are used). 
} 
\label{table:DeltaS=0}
\end{table}

The two errors quoted in Tables \ref{table:DeltaS=1} and
\ref{table:DeltaS=0} are an estimate of subleading corrections 
and the errors due to coefficients of the Gegenbauer expansion of
LCDAs. Since the predictions are made to NLO in $\alpha_s(m_b)$ but
only to LO in $1/m_b$, the largest corrections are expected to arise
from the $1/m_b$ terms. These are estimated by independently varying
the magnitudes of the leading terms proportional to
$\lambda_{u,c,t}^{(s),(d)}$ by 20\% $\sim O(\Lambda/m_b)$ and the
strong phase by $5^\circ$. This latter variation estimates the error
on the strong phase arising from the uncalculated $\alpha_s(m_b)/m_b$
or $\alpha_s^2(m_b)$ terms. A $100\%$ error 
is assigned to predictions for branching ratios in
color-suppressed tree and  QCD penguin-dominated $\Delta S=0$ decays 
where the $1/m_b$  corrections are sizable compared to the leading
results due to the hierarchy of Wilson coefficients. 
No prediction on CP asymmetries is given for these modes or for the
affected QCD penguin-dominated $\Delta S=1$ decays.

To understand better the relative sizes of different branching ratios,
it is useful to split the amplitudes for the semi-inclusive 
decays according to the CKM elements. Using the unitarity of the CKM
matrix $\lambda_t^{(q)}=-\lambda_u^{(q)} - \lambda_c^{(q)}$, 
the amplitude can be rewritten in terms of the ``tree'' amplitude
$T_{B\to   M X}$ and the ``penguin'' amplitude $P_{B\to M X}$ as
\beq\label{Asplit}
A(B\to M X)=\lambda_u^{(q)} T_{B\to M X} + \lambda_c^{(q)} P_{B\to M
  X}, 
\eeq
with $q=d,s$ for $\Delta S=0,1$ decays respectively. The ``tree'' 
amplitudes receive contributions from $O_{1,2}^u$ in
Eq.~\eqref{weakhamiltonian}, the ``penguin'' amplitudes from
$O_{1,2}^c$ (charming penguins), while the QCD and electroweak  
penguin operators contribute to both amplitudes. The
combinations of  the CKM elements exhibit the following hierarchy  
\beq
\lambda_c^{(s)}\sim\lambda_C^2,\quad
\lambda_{u,c}^{(d)}\sim\lambda_C^3,\quad 
\lambda_u^{(s)}\sim\lambda_C^4 ,
\eeq
where $\lambda_C=0.23$ is the 
Cabibbo angle. In $\Delta S=0$ decays, the two CKM factors in
Eq.~\eqref{Asplit} are of comparable size. 
In $\Delta S=1$ decays, on the other hand, there is a 
hierarchy between the two terms in Eq.~\eqref{Asplit} since
$|\lambda_u^{(s)}/\lambda_c^{(s)}|\sim \lambda_C^2$. To first order in
this small ratio, the quantity 
\beq
A_{CP}^{\Delta S=1}(B\to M X)=-2
\mathrm{Im}\left(\frac{\lambda_u^{(s)}}{\lambda_c^{(s)}}\right)
\mathrm{Im}\left(\frac{T_{B\to M X}}{P_{B\to M X}}\right) ,
\eeq
with $-2
\mathrm{Im} (\lambda_u^{(s)}/\lambda_c^{(s)})=0.037$, 
which sets a typical size of the CP asymmetries. The size of the
direct CP asymmetries also crucially depends on the ratio of ``tree''
over ``penguin'' amplitudes, as can be seen in Table
\ref{table:DeltaS=1}. This can be estimated from the sizes of the
Wilson coefficients at $\mu=m_b$ (convoluted with the asymptotic form
of LCDA) that are given in Table \ref{table:Wilson}.  
The modes that receive contributions from the operator ${\cal
  O}_{1}^u$, $\overline B_s^0\to K^{(*)-} X_{\bar s}^+ $
and $\overline B^0\to K^{(*)-} X^+ $,  have $T_{B\to M
  X}>P_{B\to M X}$ and thus have 
larger CP asymmetries. The rest of the modes listed in Table
\ref{table:DeltaS=1} do not receive these large tree contributions 
and thus have smaller CP asymmetries.  
\begin{table}
\begin{tabular}{cccccccccc}
\hline\hline
&$~~\C_{1}^u~~$ & $~~\C_{2}^u~~$ &
$~~\C_{1}^c~~$ & $~~\C_{2}^c~$ &
$~~\C_{3}~~$ & $~~\C_{4}^u~~$
&$~~\C_{4}^c~~$ &$~~\C_{5}~~$
&$~~\C_{6}~~$ 
\\\hline
{\rm Abs} & $0.89 $ & $0.080 $ & $0.0011 $ & $0.012 $ &$0.00037 $ &
$0.029 $& $0.032 $ & $0.010$ & $0.0061$ \\\hline 
{\rm Arg} & ${ 0.9^\circ}$ & ${- 99^\circ}$ & ${ 79^\circ}$ &
${181^\circ}$ &${ 7.5^\circ}$ & ${-150^\circ}$ & ${- 163^\circ}$ &
$14^\circ$ & $-151^\circ$\\ 
\hline\hline 
\end{tabular}\\[4mm]
\caption{\baselineskip 3.0ex 
The magnitudes and strong phases of the Wilson coefficients
at $\mu=m_b$ (using the notation
$\C_{3,5,6}=\C^u_{3,5,6}=\C^c_{3,5,6}$) convoluted with the asymptotic
form of the LCDA $\phi=6u \bar u$. }\label{table:Wilson} 
\end{table}
Note that the direct CP asymmetries are nonzero only if the two
interfering amplitudes in Eq.~\eqref{Asplit} have different strong
phases. In the $\Delta S=0$ decays $\overline B^0\to\phi
X^0$ and $B^-\to\phi X^-$, the two amplitudes $T_{B\to M X},P_{B\to M
  X}$ are the same at LO in $1/m_b$ and the CP asymmetries vanish. This
may change at higher orders, but  no prediction for
$A_{CP}$ for these modes is given in Table~\ref{table:DeltaS=0}. 

For color-suppressed two-body decays, the leading order tree amplitude
in $\SI$ comes from $O_{2}$.  However, when matching onto $\SII$, the
hard-spectator contribution from $O_{1}$ can compete with the leading
order term. For semi-inclusive decays considered in this paper, in
which the spectator quark does not enter the outgoing meson $M$, there
are no hard-spectator interactions.  Thus, due to the hierarchy
of the Wilson coefficients (using values in Table~\ref{table:Wilson}) 
\beq
\Big|\int_0^1 du\; 6 u\bar u \, \C_{2}^u(u)\Big|\sim 0.10\;
\Big|\int_0^1 du\; 6 u \bar u \, \C_{1}^u(u)\Big|,  
\eeq
numerically \cite{Bauer:2004tj,Williamson:2006hb}, semi-inclusive tree 
amplitudes receiving contributions from 
${\cal O}_2$ are smaller than the tree amplitudes due to ${\cal O}_1$.
The color-suppressed tree decays are then more sensitive to $1/m_b$
corrections, as discussed in the previous section. These may be
especially important for the decay $\overline B_s^0\to \omega X_s^0$
in which a cancellation between different contributions occurs for
central values of input parameters. A strong dependence of the
predictions on $a_{2\omega}$ is thus found with ${\rm Br}(
\overline{B}_s^0\to \omega X_{\bar s}^0)/{\rm Br}( \overline{B}_s^0\to
\gamma X_s^0)\in [0.003,0.027]$ for $a_{2\omega}\in [-0.3,0.3]$. 
Sizable $1/m_b$ corrections are expected in all the modes without
the charming penguin contributions: 
$\overline{B} \rightarrow \phi X^0$, $B^- \rightarrow \phi X^-$, and
$\overline{B}_s^0 \rightarrow MX_{s\bar s}^0$ ($M=\pi^0, \rho^0, \omega$). 
These decay modes  are a good experimental
source to analyze the corrections at order $1/m_b$.
A testing ground for the charming penguins are the processes in which
the tree-level amplitudes are not suppressed, and there is a charming
penguin. They correspond to processes with ${\cal C}_1^p$ and ${\cal
  C}_4^p$ in Table~\ref{table:TM}. 

In Tables \ref{table:DeltaS=1} and \ref{table:DeltaS=0} we also give
the experimental lower bounds on the predicted semi-inclusive
branching ratios. These were obtained by summing over the  
already measured two-body decays and normalizing them to ${\rm
  Br}(b\to s\gamma)=(317\pm23)\times 10^{-6}$ for ${E_\gamma >2.0~{\rm
    GeV}}$. The two-body channels for which only upper bounds are
known were not used in the estimate, nor were the decays to more than
two hadrons in the final state.

Experimentally, the semi-inclusive hadronic decays can be measured
either by summing over exclusive decays or by performing a truly
inclusive measurement where only the flavor and charge of the decaying
$B$ meson and of the isolated energetic light meson $M$ are
tagged. For these measurements a first step might be made by making an
even more inclusive measurement where only the flavor, but not the
charge of the initial $B$ meson is tagged. Theoretically simple
predictions can be made for $B^-/\overline B^0\to K_{S,L} X_s$,
$B^-/\overline B^0\to K^{*0} X_s$ and $B^-/\overline B^0\to\phi X_s$
decays, where $B^-/\overline B^0$ denotes a sum over  
the decay widths, $\Gamma(B^-\to M X)+\Gamma(\overline B^0\to M X)$. 
Using isospin symmetry, the following relations hold in the endpoint
region due to factorization at leading order in $1/m_b$:
\begin{alignat}{2}
{\rm Br}(B^-\to K_{S,L} X_s^-)&={\rm Br}(\overline B^0\to K_{S,L}
X_s^0), \quad& 
{\rm Br}(B^-\to K^{*0} X_s^-)&={\rm Br}(\overline B^0\to K^{*0} X_s^0),
\nonumber \\
{\rm Br}(B^-\to \phi X_s^-)&={\rm Br}(\overline B^0\to\phi
X_s^0),& 
{\rm Br}(B^-\to \phi X^-)&={\rm Br}(\overline B^0\to\phi X^0),
\end{alignat}
so that
\beq
\begin{split}
{\rm Br}(B^-/\overline B^0\to K_{S,L} X_s)&=\frac{1}{2}{\rm
  Br}(B^-/\overline B^0\to K^{0} X_s)=\frac{1}{4}{\rm Br}(B^-\to K^0
X_s^-),\\ 
{\rm Br}(B^-/\overline B^0\to K^{*0} X_s)&=\frac{1}{2}{\rm Br}(B^-\to
K^{*0} X_s^-),\\ 
{\rm Br}(B^-/\overline B^0\to\phi X_{(s)} )&=\frac{1}{2}{\rm
  Br}(B^-\to \phi X_{(s)}^-).
\end{split}
\eeq
For the direct CP asymmetries of these more inclusive modes, we find
\beq
\begin{split}
A_{CP}(B^-/\overline B^0\to K_{S,L} X_s)&=A_{CP}(B^-/\overline B^0\to
K^{0} X_s)=A_{CP}(B^-\to K^0 X_s^-),\\ 
A_{CP}(B^-/\overline B^0\to K^{*0} X_s)&=A_{CP}(B^-\to K^{*0} X_s^-),\\
A_{CP}(B^-/\overline B^0\to\phi X_{(s)} )&=A_{CP}(B^-\to \phi
X_{(s)}^-). 
\end{split}
\eeq
and $A_{CP}(B^-/\overline B^0\to\phi X_{u+d+s})\simeq
A_{CP}(B^-/\overline B^0\to\phi X_{s})$. 

Furthermore, for $\overline{B}^0\to \phi X$ and $\overline{B}^0 \to
\overline K^{*0} X$ decays an even more inclusive measurement can be
made, where the strangeness content of the inclusive jet need not be
determined, simplifying the measurement. Since  ${\rm
  Br}(B^-/\overline B^0\to\phi X_s)\gg {\rm Br}(B^-/\overline
B^0\to\phi X)$, the theoretical prediction for this inclusive
measurement is ${\rm Br}(B^-/\overline B^0\to\phi X_{u+d+s})\simeq
{\rm Br}(B^-/\overline B^0\to\phi X_{s})$ valid up to corrections at
the percent level. A similar simplification occurs  in $B^-/\overline
B^0\to K^{*0} X_s$ decays, since the decays $B^-\to K^{*0} X^-$ and
$\overline B^0\to K^{*0} X^0$ are absent. Therefore the strangeness of
the inclusive jet is fixed automatically and need not be determined
experimentally. An important part of the measurement is that the
flavor of $K^{*0}$ is tagged from the decay $K^{*0}\to K^+\pi^-$. 
On the other hand in $B^-/\overline B^0\to K_{S,L} X_s$ decays, since
there are contributions with the spectator quark ending up in
$\overline K^0$ from $\overline B^0\to \overline K^{0} X^0$, the
strangeness content of the inclusive jet should be  determined from
experiment.

\section{Conclusions} \label{conclusion}
In the framework of SCET we have considered semi-inclusive, hadronic
decays $B\to X M$ in the endpoint region, where the light meson $M$
and the inclusive jet $X$ with $p_X^2\sim \Lambda m_b$ are emitted
back-to back. We have considered the decays in which the spectator
quark does not enter into the meson $M$. In SCET the four-quark
operators factorize, which allows for a systematic theoretical
treatment. After matching the effective weak Hamiltonian in full QCD
onto $\SI$, the weak interaction four-quark operators factor into the
heavy-to-light current and the $\n$-collinear current. The forward  
scattering amplitude of the heavy-to-light currents leads to a
convolution $\mathcal{S}$ of the jet function with the $B$-meson shape
function, while the matrix element of $\n$-collinear currents gives
the LCDA for the meson $M$, leading to a factorized form for the decay
rates. The two nonperturbative functions, the convolution
$\mathcal{S}$ and the LCDA, are the only nonperturbative input in the
predictions for $B\to X M$ decay rates at leading order in
$1/m_b$. Furthermore, the same convolution $\mathcal{S}$  appears in
$B\to X_s \g$ decay and drops out in the ratio of $B\to X M$ to the
$B\to X_s\g$ rate and in the prediction for direct CP asymmetries.   

This greatly reduces hadronic uncertainties, since the remaining
nonperturbative input, the LCDA, is well described by its asymptotic
form, corrections to which can be obtained from other experiments or
from QCD sum rules. The Wilson coefficients can be perturbatively
computed and are then evolved to the scale $\mu_0 \sim\sqrt{\Lambda
  m_b}$ using the NLL expressions. In the ratios
the multiplicative RG evolution factors almost cancel. The
predictions for branching ratios and CP asymmetries are then given at
NLO in $\alpha_s(m_b)$ and at LO in $1/m_b$ and are collected in  
Tables \ref{table:DeltaS=1} and \ref{table:DeltaS=0}. Numerical values
are given in the limit of perturbative charming penguin due to a lack
of experimental data, while the formalism used is extended to the case
of nonperturbative charming penguins. To leading order in SCET,
the charming penguin contribution factorizes and is
given by a universal nonperturbative function ${\mathcal F}_{c\bc}$
describing the usoft interactions between the on-shell charm pair and
the bound state of the $b$ quark. In particular, ${\mathcal F}_{c\bc}$
does not depend on the final meson $M$ or the flavor content of the
inclusive jet, but only on the flavor of initial $B$ meson.  

We have also estimated subleading corrections and identify potentially
large subleading usoft contributions coming from the $\n$-collinear
sector giving rise to color-octet operators. These contributions can
be of appreciable size, compared with the leading contributions when
the leading contributions are suppressed by Wilson coefficients. This,
for instance, happens for color-suppressed tree decays and QCD penguin
(not charming penguin) dominated decays. Other contributions such as
the $(S-P)\otimes (S+P)$ operators that have been argued to be large
in exclusive $B$ decays, on the other hand, vanish 
to first order in $1/m_b$, but are present at higher
orders. Similarly, subleading corrections coming from the
heavy-to-light sector and giving subleading $B$-meson 
shape functions largely cancel in the ratio with the rate
$\overline{B} \to X_s \g$.   

In conclusion, semi-inclusive hadronic $B$ decays are a good
test field to clarify many hadronic uncertainties common to two-body
exclusive $B$ decays and the inclusive $B$ decays at the endpoint. The
factorized results provide us with a simplified view on the diverse
channels of hadronic $B$ decays  and enable us to consider them
rigorously within the framework of SCET. By investigating decays
without charming penguins, we can test whether the formalism is
working.  Then by looking at modes where the charming penguin 
can contribute, we can potentially see whether or not the charming
penguin give a large contribution to the decays.

\section*{Acknowledgments}
We thank F.~Blanc, I.~Rothstein, and J.~Smith for discussions.
J.~C.~is supported by the Korea Research Foundation Grant
KRF-2005-015-C00103. C.~K.~and A.~K.~L.~are supported in part
by the National Science Foundation under Grant No.~PHY-0244599. 
Adam Leibovich is a Cottrell Scholar of Research Corporation. JZ is
supported in part by the United States Department of Energy  
under Grants No.\ DOE-ER-40682-143 and DEAC02-6CH03000.

\appendix
\section{The Wilson coefficients at NLO } \label{appA}
The matching of the weak Hamiltonian in Eq.~\eqref{hscet} from full
QCD to $\SI$ was calculated at NLO in $\alpha_s(m_b)$  
first in  Refs.~\cite{Beneke:1999br}, and then in
Ref.~\cite{Chay:2003ju}.  For the detailed matching procedure in
obtaining the Wilson coefficients, the reader is referred to
Ref.~\cite{Chay:2003ju}.  Here we translate the results to the basis 
choice of Eqs.~\eqref{siop} and \eqref{octet}. The Wilson coefficients for
operators \eqref{siop} are\footnote{Note that
  $\C_{3,5,6}^u(v)=\C_{3,5,6}^c(v)$, so we also use the notation
  $\C_{3,5,6}(v)\equiv \C_{3,5,6}^p(v)$.} 
\begin{align} 
\begin{split}
&\C_{1,2}^p(v) = \delta_{up}\Bigl[ C_{1,2}+\frac{C_{2,1}}{N}
+\frac{\alpha_s C_F}{4\pi}\Bigl(C_{1,2}\mathcal{K}+\frac{C_{2,1}}{N}
\mathcal{F}\Bigr)\Bigr]+\frac{3}{2}\Bigl[C_{10,9}+\frac{C_{9,10}}{N}
+\\ 
&\qquad\qquad+\frac{\alpha_s C_F}{4\pi}\Bigl( C_{10,9} \mathcal{K}
+\frac{C_{9,10}}{N} \mathcal{F} \Bigr) \Bigr],  
\end{split}
\\
&\C_3^p(v) =\frac{3}{2} \Bigl[ C_{7}+\frac{C_{8}}{N}
+\frac{\alpha_s C_F}{4\pi} \Bigl(C_{7}\mathcal{K}  
+\frac{C_{8}}{N} \tilde{\mathcal{F}} \Bigr)\Bigr], 
\end{align}
\begin{align}
\begin{split}
&\C_{4,5}^p(v) =  C_{4,3}+\frac{C_{3,4}}{N} +\frac{\alpha_s
  C_F}{4\pi}\Bigl(C_{4,3}\mathcal{K}+\frac{C_{3,4}}{N}
\mathcal{F}\Bigr)-\frac{1}{2}\Bigl[C_{10,9}+\frac{C_{9,10}}{N} +\\ 
&\qquad\qquad+\frac{\alpha_s C_F}{4\pi}\Bigl( C_{10,9} \mathcal{K}
+\frac{C_{9,10}}{N} \mathcal{F} \Bigr)
\Bigr]+\frac{\alpha_s}{4\pi}\frac{C_F}{N}\big\{L_p,0\big\}, 
\end{split}
\\
\begin{split}
&\C_{6}^p(v) = C_{5}+\frac{C_{6}}{N}
+\frac{\alpha_s C_F}{4\pi} \Bigl(C_5\mathcal{K}  
+\frac{C_6}{N} \tilde{\mathcal{F}} \Bigr)-\frac{1}{2} \Bigl[
C_{7}+\frac{C_{8}}{N}  
+\frac{\alpha_s C_F}{4\pi} \Bigl(C_{7}\mathcal{K}  
+\frac{C_{8}}{N} \tilde{\mathcal{F}}  \Bigr)\Bigr],
\end{split}
\end{align}
with the shorthand notation $\mathcal{K}(v) = -6-{\pi^2}/{12}$ and
\begin{align} 
\begin{split}
\mathcal{F}(v) =& -24
-\frac{\pi^2}{12}-3i\pi+3\Bigl(1-\frac{v}{\bar v}\Bigr) \ln v +
\Bigl[(1+2i\pi) \ln^2  v 
-\frac{1-3v}{1-v} \ln v \\
&-2\ln^2 v 
- 2\mathrm{Li}_2 (1-v) - ( v \leftrightarrow \vv)\Bigr], 
\end{split}
\\ 
\tilde{\mathcal{F}}(v) =& \mathcal{F}(v)+6i \pi+24+3(\bar
v-v)\Bigl[\frac{\ln\bar v}{v}-\frac{\ln v}{\bar v}\Bigr].    
\end{align}

The contribution of a fermion loop and the gluonic operator to
$\C_{4}^p(v)$ is given as
\beq
\begin{split} 
L_p =&  \frac{2}{3}\Bigl(C_1 + 2C_3
+ 5C_4 -C_9+\frac{C_{10}}{2}
\Bigr)-\Bigl(C_3+3C_4+3C_6-\frac{C_9}{2}\Bigr) G(0)  \\ 
&-\Bigl(C_4+C_6+C_8+C_{10}\Bigr)
G(z_c)-\Bigl(C_3+C_4+C_6-\frac{1}{2}\big(C_8+C_9+C_{10}\big)\Bigr)
G(1)   \\ 
&-\frac{2}{\vv} \Bigl(C_5+C_g-\frac{1}{2}C_7 \Bigr)
-C_1\big(\delta_{up}G(0)+\delta_{cp}G(z_c)\big), 
\end{split}
\eeq       
where $z_f = m_f^2/m_b^2$ and
\begin{equation} 
G(z_f,v) = -4 \int^1_0 dw~ w(1-w) \ln\Bigl[z_f-w(1-w)\vv-i\epsilon
\Bigr]. 
\end{equation}    
The Wilson coefficients for the octet operators in Eq.~\eqref{octet}
are 
\begin{align} 
\begin{split}
&\overline{\C}_{1,2}^p(v) = \Big(2\delta_{up} C_{2,1}
+3C_{9,10}\Big)\Bigl[1+\frac{\alpha_s }{4\pi}\big(C_F\mathcal{F}-N
\mathcal{G}\big)\Bigr]+\Big(2\delta_{up} C_{1,2}
+3C_{10,9}\Big)\frac{\alpha_s }{4\pi}\mathcal{H},  
\end{split}
\\
&\overline{\C}_{3}^p(v) = 3C_{8}\Bigl[1+\frac{\alpha_s
}{4\pi}\big(C_F\tilde{\mathcal{F}}-N
\tilde{\mathcal{G}}\big)\Bigr]+3C_{7}\frac{\alpha_s
}{4\pi}\tilde{\mathcal{H}},  
\\
\begin{split}
&\overline{\C}_{4,5}^p(v) = \Big(2 C_{3,4}
-C_{9,10}\Big)\Bigl[1+\frac{\alpha_s }{4\pi}\big(C_F\mathcal{F}-N
\mathcal{G}\big)\Bigr]+\Big(2C_{4,3} -C_{10,9}\Big)\frac{\alpha_s
}{4\pi}\mathcal{H}\\ 
&\qquad\qquad-\frac{\alpha_s }{4\pi}\frac{C_F}{N}\big\{L_p,0\big\}, 
\end{split}
\\
\begin{split}
&\overline{\C}_{6}^p(v) = \Big(2 C_{6}
-C_{8}\Big)\Bigl[1+\frac{\alpha_s }{4\pi}\big(C_F\tilde{\mathcal{F}}-N
\tilde{\mathcal{G}}\big)\Bigr]+\Big(2C_{5} -C_{7}\Big)\frac{\alpha_s
}{4\pi}\tilde{\mathcal{H}}, 
\end{split}
\end{align}
where
\begin{align}
\mathcal{G}=&\frac{1}{2}\Bigl[-10 +\ln^2  v - \frac{2}{v} \ln\vv  +
\ln^2 \vv -2\mathrm{Li}_2\Bigl(-\frac{v}{\vv}\Bigr)- \frac{7}{6}\pi^2
-2i\pi \ln v \Bigr], 
\\
\mathcal{H}=& \frac{1}{2}\Bigl[-18+(2-3v)\Bigl(\frac{\ln
  v}{\vv}-\frac{\ln\vv}{v}\Bigr)   
+ 2\mathrm{Li}_2\Bigl(-\frac{\vv}{v}\Bigr) -
2\mathrm{Li}_2\Bigl(-\frac{v}{\vv}\Bigr)-3i\pi \Bigr], 
\\
\tilde{\mathcal{G}} =&\frac{1}{2}\Bigl[2 +\ln^2  v -3\ln v +
\frac{1-3v}{v} \ln\vv   
+ \ln^2\vv-2\mathrm{Li}_2\Bigl(-\frac{v}{\vv}\Bigr)  -
\frac{7}{6}\pi^2+i\pi (3- 2\ln v ) \Bigr],\\ 
\tilde{\mathcal{H}} =& \frac{1}{2}\Bigl[6 -(1-3v)\Bigl(\frac{\ln v}{\vv}
-\frac{\ln\vv}{v}\Bigr) + 2\mathrm{Li}_2\Bigl(-\frac{\vv}{v}\Bigr) -
2\mathrm{Li}_2\Bigl(-\frac{v}{\vv}\Bigr)+3i\pi \Bigr].  
\end{align}

\section{Nonperturbative charming penguin in the heavy quark
  limit}\label{appB} 
In this Appendix we show that in the heavy quark limit,
$m_b, m_c\to\infty$ with $r = m_c/m_b$ fixed, the
charming penguin contributions to the decay rates factorize in SCET
into hard, jet, collinear, and soft parts at LO in $1/m_{c,b}$.
A typical charming penguin contribution is shown in Fig.~\ref{charm}. 
When the momentum transfer in the gluon is close
to $4m_c^2$, the intermediate charm quark pair is nearly
on-shell and can have usoft interactions.
In the $B$ meson rest frame, the velocity of the $b$ quark can be
written as $v^{\mu} = (n^{\mu} + \n^{\mu})/2$ with $n^{\mu} =
(1,0,0,1)$ and $\n^{\mu} = (1,0,0,-1)$. In this frame, 
the on-shell charm quark pair has momentum $2m_c v_{\bar
  cc}^{\mu} + k^{\mu}$, where $k^{\mu}\sim \Lambda$ is the residual
momentum, while $v_{\bar  cc}^{\mu}$ is the velocity of the charm
quark pair with $v_{\bar  cc}^2 =1$. It is given by
\begin{equation} 
v_{\bar cc}^{\mu} =  \frac{1}{2r}  \frac{n^{\mu}}{2} + 
\frac{\bar{u}x_M}{2r} \frac{\n^{\mu}}{2}= \beta \frac{n^{\mu}}{2} +
\frac{1}{\beta}  \frac{\overline{n}^{\mu}}{2},
\label{cv1}
\end{equation}  
where $4r^2 = \bar{u} x_M$, with $x_M$ close to 1 and $\bar{u} = 1-u$. 
 
The charm quark pair annihilates into a gluon with off-shellness
of order $4m_c^2 \sim m_b^2$. Integrating out the intermediate
off-shell gluon gives a four-quark operator at leading order in
$1/m_c$ 
\begin{equation} 
{\cal O}_{c\bar{c}n\bar{n}} 
= \sum_q (\overline{q}_{n,\w} \gamma^{\mu} T^a  
 q_{\n,\bw_1})~(\overline{c}_{-v_{\bar cc}} \gamma_{\mu} T^a 
 c_{v_{\bar cc}}),
\end{equation}
where the charm quarks are treated as heavy. The  collinear quark
fields $q_n$ and $q_{\bar{n}}$ are defined as 
\beq
\begin{split}
q_{\n,\bw_1}&=\Bigl[\delta (\bw_1-n\cdot \mP) W_{\n}^{\dagger} 
 \xi^{q}_{\n}\Bigr], \qquad
\overline{q}_{n,\w}=\Bigl[\overline{\xi}_n^{q} W_n \delta(\w - \n\cdot
\mP^{\dagger}) \Bigr], 
\end{split}
\eeq
where $W_n$ ($W_{\bar{n}}$) is the collinear Wilson line in the $n$
($\overline{n}$) direction from integrating out off-shell heavy charm
quarks. Note that these collinear Wilson lines are the same as those
from the heavy $b$ quarks in Eq.~\eqref{qn}. This is a manifestation
of Type-III reparameterization invariance
\cite{Chay:2002vy,Manohar:2002fd}, which states that the SCET
Lagrangian and the collinear Wilson lines are invariant under $n_{\mu}
\to n_{\mu}/\beta $ and $\n_{\mu} \to \beta \n_{\mu}$ with $\beta$
close to 1. For example, the collinear Wilson line $W_n$ is invariant
under this transformation as  
\begin{equation}  
W_{n'} \Biggl( \frac{\n'\cdot A}{\n'\cdot \mP} \Biggr)
=W_{n'} \Biggl( \frac{\beta \n\cdot A}{\beta \n\cdot \mP} \Biggr)
= W_{n} \Biggl( \frac{\n\cdot A}{\n\cdot \mP} \Biggr),
\end{equation}
which also holds for $W_{\bar{n}}$. This corresponds to the Lorentz
invariance under a boost with $\beta=m_b/(2m_c)$ in the $z$ direction,
corresponding to transforming to the $B$ meson rest frame. 
The usoft interactions can be decoupled from collinear
interactions by introducing the usoft Wilson lines 
$Y_n$ and $Y_{\bar{n}}$  and redefining the collinear fields
\cite{Bauer:2001yt}. This gives 
\begin{equation} 
{\cal O}_{c\bar{c}n\bar{n}} 
= \sum_q (\overline{q}_{n,\w} \gamma^{\mu} T^a  
 q_{\n,\bw_1})~(\overline{c}_{-v_{\bar cc}} Y_{\n} \gamma_{\mu} T^a
 Y_n^{\dagger} 
 c_{v_{\bar cc}}),
\label{occ2}
\end{equation}
where the collinear fields from now on will denote the redefined fields.
The operator ${\cal O}_{c\bar{c}n\bar{n}}$ satisfies 
gauge invariance in SCET \cite{Bauer:2001yt,Bauer:2003mg}, and the
subleading corrections to this operator will be of order
$\Lambda/m_{c,b}$.

Let us next discuss the form of the nonperturbative charming penguin
contributions that arise from the time ordered product 
of ${\cal O}_{c\bar{c}n\bar{n}} $ with the operators in the weak
Hamiltonian. We work out the details for
the operator $O_1^c = 4 (\bar{q} \gamma^{\nu} P_L c) 
(\bar{c} \gamma_{\nu} P_L b)$ ($q=d,s$) that matches onto the SCET
operator 
\beq \label{oqb}
{\cal O}_{qbc\bar{c}}(\bw_2) =
4 \Bigl[(\overline{q}_{\n,\bw_2})_{\beta} 
\gamma^{\nu} P_L (Y_n^{\dagger} b_v )_{\alpha} \Bigr] 
\Bigl[ ( \overline{c}_{v_{\bar cc}} Y_n^{\dagger} )_{\alpha} 
\gamma_{\nu} P_L (Y_{\n}^{\dagger} c_{-v_{\bar cc}})_{\beta} \Bigr], 
\eeq 
where $\alpha$, $\beta$ are color indices. The treatment of other
operators is similar. The matrix element for the
contribution of weak  operator $O_1^c$ 
in Fig.~\ref{charm} is then 
\beq 
\begin{split}
\label{hcc}
\langle M X | H_W^{c\bar{c}} | B \rangle
&=\frac{G_F}{\sqrt{2}} \lambda_c^{(q)} ~i \int d\w d\bw_1 d\bw_2\int
d^4 y  \sum_{\tilde{p}_n, \tilde{p}_{\bar{n}}}
e^{i(-2m_c v_{\bar cc}+ \tilde{p}_n - \tilde{p}_{\bar{n}})\cdot y} 
\\
&\times  
H_{c\bc n\n} (\w, \bw_1) H_{qbc\bc}(\bw_2) \;
\langle M X |\mathrm{T} {\cal O}_{c\bar{c}n\n}(\w,\bw_1, y)
{\cal O}_{qbc\bar{c}}(\bw_2,0) | B \rangle,
\end{split}
\eeq
where $H_{c\bc n\n} = -4\pi\alpha_s/(\omega\bar\omega_1)$ and
$H_{qbc\bc}=1+{\cal O}(\alpha_s)$ are the Wilson coefficients of the  
operators ${\cal O}_{c\bar{c}n\n}$  and
${\cal O}_{qbc\bar{c}}$ in Eqs.~(\ref{occ2}) and
(\ref{oqb}). 

Using the factorization of $\n$-collinear quarks from usoft and
$n$-collinear degrees of freedom the matrix element \eqref{hcc} 
can be rewritten as 
\begin{equation} \label{hcc1}
\frac{G_F}{\sqrt{2}N} \lambda_c^{(q)} c_q^{BM} h_M \int^1_0 du\,
 ~\delta \Bigl(\bar{u}-\frac{4r^2}{x_M}\Bigr)\phi_M (u) H(u,x_M)  
~\langle X | \mathcal{Q}_{c\bar{c}} | B \rangle, 
\end{equation} 
where
\begin{eqnarray}
\langle X | \mathcal{Q}_{c\bar{c}} | B \rangle&=& -4 i\int
d^4y \langle X 
|\mathrm{T}   
~\overline{q}_n (y)   \gamma^{\mu} \Gamma_M
\gamma^{\nu} P_L T^a (Y_{\n}^{\dagger} c_{-v_{\bar cc}} )(0) \nonumber\\
&&\times (\overline{c}_{-v_{\bar cc}} Y_{\n} \gamma_{\mu} T^a
Y_n^{\dagger} 
c_{v_{\bar cc}} )(y) (\bar{c}_{v_{\bar cc}} Y_n 
\gamma_{\nu} P_L Y_n^{\dagger} b_v) (0) | B\label{Qccbar}
\rangle.  
\end{eqnarray}
The delta function in \eqref{hcc1} is obtained from the exponent of the 
label momenta in Eq.~(\ref{hcc}) using $\n\cdot p = m_b$, 
$n\cdot p_M = x_M m_b$. The hard kernel is $H(u,x_M) = 4\pi
\alpha_s(2m_c) C_1/ (\bar{u}x_M m_b^2) + {\cal O}(\alpha_s^2)$. 
In obtaining (\ref{hcc1}) the relation 
\begin{equation} 
\langle M | (q_{\n} )_a \big[\overline{q}_{\n} \;\delta\big(u-{n\cdot
  \mP^{\dagger}}/{n\cdot p_M}\big)\big]_b| 0 \rangle 
= - h_M (\Gamma_M)_{ab}~\phi_M (u)
\label{wave}
\end{equation}
was used, with $M= P, L, T$ denoting pseudoscalar, longitudinally, and 
transversely polarized vector mesons, respectively. The product $h_M
\Gamma_M$ is 
\begin{equation} 
 h_M \Gamma_M = \! \frac{n\cdot p_M}{8} \left\{ \begin{array}{l} 
\displaystyle{i}f_P  {\fms{\n}}
 \gamma_5, \quad\; (M=P), \\ 
\displaystyle if_V  {\fms{\n}}, \qquad\; (M=L),
\\  
\displaystyle f_V^{\perp} 
\eta^{*\alpha}_{\perp} \gamma_{\alpha}^{\perp}{\fms{\n}} \;\;
(M=T).  
\end{array} \right.
\label{vtom}
\end{equation} 
The coefficients $c_q^{BM}$ describe the flavor content of the meson
$M$ and  are $\sqrt{2} c_d^{B_s M} =(-1,-1,1)$ for $\overline{B}_s^0
\rightarrow (\pi^0, \rho^0, \omega)$, while $c_q^{BM} =1$ for other 
decays.

\begin{figure}[t]
\begin{center}
\epsfig{file=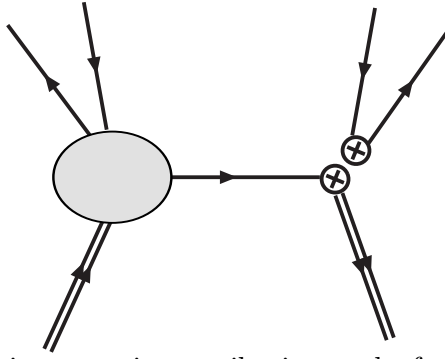, width=6cm}
\end{center}
\vspace{-1.0cm}
\caption{\baselineskip 3.0ex 
Nonperturbative charming penguin contribution to the forward
scattering amplitude. The blob  is the nonperturbative charm 
contribution and the mirror image is omitted.}
\label{ch}
\end{figure}

In order to obtain the corrections from nonperturbative charming
penguin to the inclusive decay rates, the optical theorem is used in a
similar way as in section~\ref{fac}. To first order in
$\alpha_s(2m_c)$ only  the time-ordered product shown in Fig.~\ref{ch}, 
\beq \label{tcc}
\begin{split}
\mathcal{T}_{c\bar{c}} =& \frac{i}{m_B} \int d^4z e^{-ip'\cdot z} 
\langle B | \mathrm{T} J_H^{\dagger} (z) \mathcal{Q}_{c\bc}
(0)| B \rangle  \\
=& 4\int d^4z\;d^4 y\; e^{im_b(1-x_M)\n\cdot z/2}
 \langle B_v| \mathrm{T}  
\big[\overline{b}_v Y_n \fms{\n} P_L q_n\big](z) \\
&\times\big[(\overline{q}_n)(y)  \gamma^{\mu} \Gamma_M
\gamma^{\nu} P_L T^a (Y_{\n}^{\dagger} c_{-v_{\bar cc}})(0)\big]
\big[\overline{c}_{-v_{\bar cc}} Y_{\n} \gamma_{\mu} T^a
Y_n^{\dagger} c_{v_{\bar cc}}\big](y) \\  
&\times\big[\bar{c}_{v_{\bar cc}} Y_n^{\dagger} \gamma_{\nu} P_L
Y_n^{\dagger} b_v \big](0) | B_v \rangle,
\end{split}
\eeq
and its hermitian conjugate are needed.
The time-ordered product $\langle B | \mathrm{T}
\mathcal{Q}_{c\bc}^{\dagger} (z) \mathcal{Q}_{c\bc} 
(0)| B \rangle$ contributes at order $\alpha_s^2(2m_c)$ and is
neglected in our discussion.  In Eq.~(\ref{tcc}), the time-ordered 
product of the $n$-collinear fields can be factored out into the  
jet function
\beq 
\langle 0| \mathrm{T} q_n (z)  \overline{q}_n (y)
|0\rangle = i\frac{\fms{n}}{2} \delta(n\cdot(z-y))\delta(z_\perp-y_{\perp})
\int \frac{dn\cdot k}{2\pi} e^{-in\cdot k \n\cdot
  (z-y)/2}  
J_{\n\cdot p}(n\cdot k + 
i\epsilon), \label{jet} 
\eeq
and $\mathcal{T}_{c\bar{c}}$ becomes
\beq 
\begin{split}
\mathcal{T}_{c\bc} =&  4\int\frac{dn\cdot k\,d\n\cdot z }{4\pi}  
i\! \int d^4 y ~e^{i[m_b(1-x_M)-n\cdot k]\n\cdot z/2} 
e^{in\cdot k \n\cdot y/2} J_{m_b} (n\cdot k +i\epsilon) \\
\label{tcc1}
&\times\langle B_v|\mathrm{T}\;\overline{b}_v Y_n
(\tfrac{\n\cdot z}{2}) \nn P_L \frac{\fms{n}}{2} \gamma^{\mu}
\Gamma_M 
\gamma^{\nu} P_L T^a Y_{\n}^{\dagger} c_{-v_{\bar cc}} (0)  \\ 
&\times \overline{c}_{-v_{\bar cc}} Y_{\n} \gamma_{\mu} T^a
Y_n^{\dagger} c_{v_{\bar cc}} (y) 
\cdot \bar{c}_{v_{\bar cc}} Y_n^{\dagger} 
\gamma_{\nu} P_L Y_n^{\dagger} b_v  (0) | B_v \rangle. 
\end{split}
\eeq

If the meson $M$ is a transversely polarized vector 
meson, $\Gamma_M= \gamma_{\perp}^{\alpha} \fms{\n}/2$, and
$\mathcal{T}_{c\bc}$ vanishes because of the spin symmetry.  
Eq.~(\ref{tcc1}) thus implies
that charming penguin effects could give a contribution to $B\to V_T
X$ decays only at subleading orders in $\Lambda/m_c$ and/or
$\alpha_s(2m_c)$. We expect that a similar conclusion holds also for
the two-body nonleptonic exclusive decays.  
There large transverse polarization fractions, $R_T \sim 0.5$, have
been measured in $\Delta S=1$ 
$B\to VV$ decays (such as $B\to \phi K^*$) that can be charming
penguin dominated \cite{Aubert:2003mm,Aubert:2004xc,Abe:2004ku}. This
may signal substantial $1/m_c$ corrections. 
In naive factorization the transverse component on the contrary is
expected to be suppressed by $\mathcal{O}(m_V^2/m_B^2)$ due to a 
spin flip. In order to explain this large transverse rate, several
possibilities of enhanced higher-order contributions in $1/m_b$ were
suggested \cite{rescattering,Kagan:2004uw,Li}. The long-distance  
charming penguin at leading order has also been proposed to contribute
to large $R_T$ \cite{Bauer:2004tj}.

For pseudoscalar or longitudinally polarized vector meson, on the
other hand, the nonperturbative charming penguin contribution is
\beq
\begin{split} 
\mathcal{T}_{c\bc} =&  8\int\frac{dn\cdot k\, d\n\cdot z}{4\pi}  
i\! \int d^4 y ~e^{i[m_b(1-x_M)-n\cdot k]\n\cdot z/2} 
e^{+in\cdot k \n\cdot y/2} J_{m_b} (n\cdot k +i\epsilon)   \\
\label{tcc2}
&\times\langle B_v|\mathrm{T}\overline{b}_v Y_n 
(\tfrac{\n\cdot z}{2})
\gamma_{\perp}^{\mu} \nn 
\gamma^{\nu} P_L T^a Y_{\n}^{\dagger} c_{-v_{\bar cc}} (0)  \\
&\times \overline{c}_{-v_{\bar cc}} Y_{\n} \gamma^{\perp}_{\mu} 
T^a Y_n^{\dagger} c_{v_{\bar cc}} (y)
\cdot \bar{c}_{v_{\bar cc}} Y_n 
\gamma_{\nu} P_L Y_n^{\dagger} b_v  (0) | B_v \rangle. 
\end{split} 
\eeq
The factorization in $\mathcal{T}_{c\bc}$ is more apparent if we 
rewrite it in a more compact form as 
\begin{eqnarray} \label{tcc3}
\mathcal{T}_{c\bc} &=& -2 \int^{\overline{\Lambda}}_{-m_b(1-x_M)} d
n\cdot l\;  f_{c\bc}^{(1)}\big(m_b(1-x_M) +n\cdot l, n\cdot l\big)\; 
J_{m_b} (n\cdot k +i\epsilon)  
\end{eqnarray}    
where we have introduced a new, in general complex, nonperturbative
function $f_{c\bar{c}}^{(1)}$ 
\beq\label{fcc}
\begin{split}
\int ^{\overline{\Lambda}}_{-m_b(1-x_M)} 
&dn\cdot l~ e^{in\cdot l \n\cdot z/2} f_{c\bc}^{(1)} (n\cdot k, n\cdot l)=\\
& -2i\!\int d^4 y~e^{in\cdot k \n\cdot y/2}\langle 
B_v|\mathrm{T} \big[(\overline{b}_v Y_n)(\tfrac{\n\cdot z}{2})
\gamma_{\perp}^{\mu} \fms\n 
\gamma^{\nu} P_L T^a (Y_{\n}^{\dagger} c_{-v_{\bar cc}})(0)\big]  \\ 
&\times \big[\overline{c}_{-v_{\bar cc}}
Y_{\n} \gamma^{\perp}_{\mu}  
T^a Y_n^{\dagger} c_{v_{\bar cc}}\big](y)
\big[\bar{c}_{v_{\bar cc}} Y_n^{\dagger} 
\gamma_{\nu} P_L Y_n^{\dagger} b_v \big] (0) | B_v \rangle.
\end{split}
\eeq
The integration over $n\cdot l$ can be interpreted as the integration
over soft fluctuations of $b_v$. Taking the discontinuity of the jet
function in $\mathcal{T}_{c\bc}$ we finally obtain 
\beq
\label{Fcc}
\mathcal{F}_{c\bc}^{(1)} = 2 \int^{\overline{\Lambda}}_{-m_b(1-x_M)} d
n\cdot l  \;
f_{c\bc}^{(1)}\big(m_b(1-x_M) +n\cdot l, n\cdot l\big)\;
\Big[-\frac{1}{\pi} \mathrm{Im}\ J_{m_b} (n\cdot k +i\epsilon)\Big].
\eeq
The jet function can be systematically computed in powers of $\alpha_s
(\sqrt{\Lambda m_b})$. Instead of pursuing this option, we can treat
the convolution of 
jet function and $f_{c\bc}^{(1)}$ as a nonperturbative function to be
determined from experiment. 
The nonperturbative charming penguin contribution to the decay rate
corresponding to a sum of Fig.~\ref{ch} and its mirror image is then 
\beq\label{gcc}
\begin{split}
\frac{d\Gamma^{(1)}_{c\bc}(B \to MX)}{dE_M}  =& 16\pi^2
\alpha_s (2m_c)\frac{G_F^2}{N} \frac{f_M^2 E_M^3}{16\pi^2} 
\frac{1}{8r^2m_b^2}~\phi_M (1-{4r^2}/{x_M})    \\ 
&\times  \lambda_c^{(q)}c_q^{BM} C_1 (m_b) \cdot 2\mathrm{Re}
\Bigl[\T_{M}^{(q)*}(m_b)  \mathcal{F}_{c\bc}^{(1)} (x_M)\Bigr].
\end{split}
\eeq
If we include all the possible contributions from the four-quark
operators, the nonperturbative charming penguin contribution to the
decay rate at leading order in $1/m_{c,b}$ and $\alpha_s(2m_c)$ is
written as
\begin{equation}  \label{facc}
\frac{d\Gamma_{\bar{c}c}(B\to X M)}{dE_M} 
= \frac{G_F^2}{8\pi} f_M^2 m_b^2 x_M^3 \alpha_s(2m_c) \lambda_c^{(q)}
c_q^{BM} \phi_M (1-{4r^2}/{x_M}) 
\cdot 2 \mathrm{Re} \T_M^{(q)*} \mathcal{F}_{c\bc},
\end{equation}  
where the hard coefficients $\T_M^{(q)}$ are listed in Table
\ref{table:TM}, while 
\begin{eqnarray}\label{Fccfull}
 \mathcal{F}_{c\bc} &=& \frac{\pi}{8Nr^2 m_b}\left\{  \left[C_1(m_b)+
     C_4(m_b) + C_{10}(m_b)\right] \mathcal{F}_{c\bc}^{(1)}\right.  
 +  \left[C_2(m_b)+ C_3(m_b) + C_9(m_b)\right]
 \mathcal{F}_{c\bc}^{(2)} \nonumber  \\
&+&\lambda_u^{(q)}/\lambda_c^{(q)} \left.\left[\big(C_4(m_b) +
    C_{10}(m_b)\big) \mathcal{F}_{c\bc}^{(1)} 
  + \big(C_3(m_b) + C_9(m_b)\big) \mathcal{F}_{c\bc}^{(2)}
\right]\right\}. 
\end{eqnarray}
The nonperturbative function $\mathcal{F}_{c\bc}^{(2)}$  arises from
the weak operators with the same color structure as $O_2^c$, so that
\begin{equation}
  \mathcal{F}_{c\bc}^{(2)} = 2
  \int^{\overline{\Lambda}}_{-m_b(1-x_M)} d n\cdot l \; 
f_{c\bc}^{(2)}\big(m_b(1-x_M) +n\cdot l, n\cdot l\big)\;
\Big[-\frac{1}{\pi} \mathrm{Im}\ J_{m_b} (n\cdot k +i\epsilon)\Big],
\end{equation}
where
\beq\label{fcc2}
\begin{split}
\int ^{\overline{\Lambda}}_{-m_b(1-x_M)} &
dn\cdot l~ e^{in\cdot l \n\cdot z/2} f_{c\bc}^{(2)} (n\cdot
k, n\cdot l) =\\
& -2i\int d^4 y~ e^{+in\cdot k \n\cdot y/2} \langle 
B_v|\mathrm{T}\overline{b}_v Y_n 
(\tfrac{\n\cdot z}{2}) \gamma_{\perp}^{\mu} \fms{\overline{n}}
\gamma^{\nu} P_L T^a Y_{\n}^{\dagger} b_v (0)  \\ 
&\times \overline{c}_{-v_{\bar cc}} Y_{\n} \gamma^{\perp}_{\mu} 
T^a Y_n^{\dagger} c_{v_{\bar cc}} (y)
\cdot \bar{c}_{v_{\bar cc}} Y_n^{\dagger} 
\gamma_{\nu} P_L Y_n^{\dagger} c_{-v_{\bar cc}}  (0) | B_v \rangle.  
\end{split}
\eeq
The terms proportional to $\lambda_u^{(q)}$ in \eqref{Fccfull} are
smaller than $<2\% \ (<0.1\%)$ of the terms in the first row of
\eqref{Fccfull} for $\Delta S=0\ (\Delta S=1)$ decays and can be
safely neglected.  

The function $\mathcal{F}_{c\bc}$ is independent of the outgoing meson $M$.  
In obtaining Eq.~\eqref{facc} an expansion in $\alpha_s (2m_c)$ was
used. If the expansion does not converge one can 
still parametrize the nonperturbative charming penguins by treating
the product of $\alpha_s (2m_c)$, the LCDA, and 
$\mathcal{F}_{c\bc}$ as a new nonperturbative parameter, to be
extracted from experiment. Unlike $\mathcal{F}_{c\bc}$, however, this new
parameter depends on $M$.


\begin{thebibliography}{99}
\normalsize
\baselineskip 3.0ex 

\bibitem{Browder:1997yq}
T.~E.~Browder, A.~Datta, X.~G.~He and S.~Pakvasa,
Phys.\ Rev.\ D {\bf 57}, 6829 (1998);
  A.~Datta, X.~G.~He and S.~Pakvasa,
  Phys.\ Lett.\ B {\bf 419}, 369 (1998);
  X.~G.~He, C.~P.~Kao, J.~P.~Ma and S.~Pakvasa,
  Phys.\ Rev.\ D {\bf 66}, 097501 (2002).

\bibitem{Atwood:1997de}
  D.~Atwood and A.~Soni,
  Phys.\ Rev.\ Lett.\  {\bf 79}, 5206 (1997).

\bibitem{Kagan:1997qn}
  A.~L.~Kagan and A.~A.~Petrov,  arXiv:hep-ph/9707354.

\bibitem{He:1998se}
 X.~G.~He and G.~L.~Lin,  Phys.\ Lett.\ B {\bf 454}, 123 (1999);
 X.~G.~He, J.~P.~Ma and C.~Y.~Wu, Phys.\ Rev.\ D {\bf 63}, 094004
 (2001);  X.~G.~He, C.~Jin and J.~P.~Ma, Phys.\ Rev.\ D {\bf 64},
 014020 (2001). 

\bibitem{Calmet:1999ix}
  X.~Calmet, T.~Mannel and I.~Schwarze,  Phys.\ Rev.\ D {\bf 61},
  114004 (2000);   X.~Calmet, T.~Mannel and I.~Schwarze,
  Phys.\ Rev.\ D {\bf 62}, 096014 (2000);  X.~Calmet,
   Phys.\ Rev.\ D {\bf 62}, 014027 (2000);  X.~Calmet,
  Phys.\ Rev.\ D {\bf 62}, 016011 (2000).

\bibitem{Cheng:2001nj}
H.~Y.~Cheng and A.~Soni, Phys.\ Rev.\ D {\bf 64}, 114013 (2001).

\bibitem{Kim:2002gv}
  C.~S.~Kim, J.~Lee, S.~Oh, J.~S.~Hong, D.~Y.~Kim and H.~S.~Kim,
  Eur.\ Phys.\ J.\ C {\bf 25}, 413 (2002).

\bibitem{Eilam:2002wu}
  G.~Eilam and Y.~D.~Yang,  Phys.\ Rev.\ D {\bf 66}, 074010 (2002).

\bibitem{Soni:2005jj}
  A.~Soni and J.~Zupan,  arXiv:hep-ph/0510325.

\bibitem{Bauer:2000ew}
C.~W.~Bauer, S.~Fleming and M.~E.~Luke, Phys.\ Rev.\ D {\bf 63},
014006 (2001). 

\bibitem{Bauer:2000yr}
C.~W.~Bauer, S.~Fleming, D.~Pirjol and I.~W.~Stewart, Phys.\ Rev.\ D
{\bf 63}, 114020 (2001).

\bibitem{Bauer:2001ct}
C.~W.~Bauer and I.~W.~Stewart, Phys.\ Lett.\ B {\bf 516}, 134 (2001). 

\bibitem{Bauer:2001yt}
  C.~W.~Bauer, D.~Pirjol and I.~W.~Stewart,  Phys.\ Rev.\ D {\bf 65},
  054022 (2002).

\bibitem{Chay:2003zp}
  J.~Chay and C.~Kim,  Phys.\ Rev.\ D {\bf 68}, 071502 (2003).

\bibitem{Chay:2003ju}
  J.~Chay and C.~Kim,  Nucl.\ Phys.\ B {\bf 680}, 302 (2004).

\bibitem{work}
J.~Chay, A.~K.~Leibovich, C.~Kim, J.~Zupan, work in progress.

\bibitem{Bauer:2003pi}
  C.~W.~Bauer and A.~V.~Manohar,  Phys.\ Rev.\ D {\bf 70}, 034024
  (2004);  S.~W.~Bosch, B.~O.~Lange, M.~Neubert and G.~Paz,
  Nucl.\ Phys.\ B {\bf 699}, 335 (2004); M.~Neubert,
  Eur.\ Phys.\ J.\ C {\bf 40}, 165 (2005).


\bibitem{Ciuchini}
  M.~Ciuchini, E.~Franco, G.~Martinelli and L.~Silvestrini,
  Nucl.\ Phys.\ B {\bf 501}, 271 (1997);
  M.~Ciuchini, E.~Franco, G.~Martinelli, M.~Pierini and L.~Silvestrini,
  Phys.\ Lett.\ B {\bf 515}, 33 (2001).

\bibitem{Bauer:2004tj}
  C.~W.~Bauer, D.~Pirjol, I.~Z.~Rothstein and I.~W.~Stewart,
  Phys.\ Rev.\ D {\bf 70}, 054015 (2004);
  C.~W.~Bauer, I.~Z.~Rothstein and I.~W.~Stewart,
  arXiv:hep-ph/0510241.

\bibitem{Williamson:2006hb}
  A.~R.~Williamson and J.~Zupan,  arXiv:hep-ph/0601214.

\bibitem{Beneke:1999br}
  M.~Beneke, G.~Buchalla, M.~Neubert and C.~T.~Sachrajda,
  Phys.\ Rev.\ Lett.\  {\bf 83}, 1914 (1999);
  M.~Beneke, G.~Buchalla, M.~Neubert and C.~T.~Sachrajda,
  Nucl.\ Phys.\ B {\bf 591}, 313 (2000);
  M.~Beneke, G.~Buchalla, M.~Neubert and C.~T.~Sachrajda,
  Nucl.\ Phys.\ B {\bf 606}, 245 (2001);
  M.~Beneke and M.~Neubert,
  Nucl.\ Phys.\ B {\bf 675}, 333 (2003).

\bibitem{Bauer:2005wb}
  C.~W.~Bauer, D.~Pirjol, I.~Z.~Rothstein and I.~W.~Stewart,
   Phys.\ Rev.\ D {\bf 72}, 098502 (2005).

\bibitem{Beneke:2004bn}
  M.~Beneke, G.~Buchalla, M.~Neubert and C.~T.~Sachrajda,
  Phys.\ Rev.\ D {\bf 72}, 098501 (2005).

\bibitem{Korchemsky:1994jb}
  G.~P.~Korchemsky and G.~Sterman,
  Phys.\ Lett.\ B {\bf 340}, 96 (1994);   R.~Akhoury and
  I.~Z.~Rothstein,   Phys.\ Rev.\ D {\bf 54}, 2349 (1996).


\bibitem{Lepage:1980fj}
  G.~P.~Lepage and S.~J.~Brodsky,
   Phys.\ Rev.\ D {\bf 22}, 2157 (1980).

\bibitem{Chernyak:1983ej}
  V.~L.~Chernyak and A.~R.~Zhitnitsky,
   Phys.\ Rept.\  {\bf 112}, 173 (1984).

\bibitem{Fleming:2004rk}
  S.~Fleming and A.~K.~Leibovich,
   Phys.\ Rev.\ D {\bf 70}, 094016 (2004).
 

\bibitem{Braun:2003rp}
  V.~M.~Braun, G.~P.~Korchemsky and D.~Mueller,
   Prog.\ Part.\ Nucl.\ Phys.\  {\bf 51}, 311 (2003).

\bibitem{Chay:2005ck}
  J.~Chay, C.~Kim and A.~K.~Leibovich,
   Phys.\ Rev.\ D {\bf 72}, 014010 (2005).
 


\bibitem{Feldmann:2004mg}
  T.~Feldmann and T.~Hurth,
  JHEP {\bf 0411}, 037 (2004).

\bibitem{Bauer:2001mh}
  C.~W.~Bauer, M.~E.~Luke and T.~Mannel,
  Phys.\ Rev.\ D {\bf 68}, 094001 (2003).

\bibitem{Leibovich:2002ys}
  A.~K.~Leibovich, Z.~Ligeti and M.~B.~Wise,
  Phys.\ Lett.\ B {\bf 539}, 242 (2002).

\bibitem{Bauer:2002yu}
  C.~W.~Bauer, M.~Luke and T.~Mannel,  Phys.\ Lett.\ B {\bf 543}, 261
  (2002). 

\bibitem{Lee:2004ja}
  K.~S.~M.~Lee and I.~W.~Stewart,  Nucl.\ Phys.\ B {\bf 721}, 325
  (2005). 

\bibitem{Bosch:2004cb}
  S.~W.~Bosch, M.~Neubert and G.~Paz,  JHEP {\bf 0411}, 073 (2004).


\bibitem{Chay:2002vy}
  J.~Chay and C.~Kim,  Phys.\ Rev.\ D {\bf 65}, 114016 (2002).


\bibitem{Manohar:2002fd}
  A.~V.~Manohar, T.~Mehen, D.~Pirjol and I.~W.~Stewart,
  Phys.\ Lett.\ B {\bf 539}, 59 (2002).

\bibitem{Hardmeier:2003ig}
  A.~Hardmeier, E.~Lunghi, D.~Pirjol and D.~Wyler,
  Nucl.\ Phys.\ B {\bf 682}, 150 (2004).

\bibitem{Leibovich:2003jd}
  A.~K.~Leibovich, Z.~Ligeti and M.~B.~Wise,
  Phys.\ Lett.\ B {\bf 564}, 231 (2003).

\bibitem{Rothstein:2003wh}
  I.~Z.~Rothstein,
  Phys.\ Rev.\ D {\bf 70}, 054024 (2004).

\bibitem{Chen:2003fp}
  J.~W.~Chen and I.~W.~Stewart,  Phys.\ Rev.\ Lett.\  {\bf 92}, 202001
  (2004). 

\bibitem{Khodjamirian:2003xk}
  A.~Khodjamirian, T.~Mannel and M.~Melcher,
  Phys.\ Rev.\ D {\bf 68}, 114007 (2003);
  V.~M.~Braun and A.~Lenz,  Phys.\ Rev.\ D {\bf 70}, 074020 (2004);
  P.~Ball and R.~Zwicky,  JHEP {\bf 0602}, 034 (2006).

\bibitem{Ball:2006wn}
  P.~Ball, V.~M.~Braun and A.~Lenz,  JHEP {\bf 0605}, 004 (2006).

\bibitem{Ball:2004rg}
  P.~Ball and R.~Zwicky,  Phys.\ Rev.\ D {\bf 71}, 014029 (2005).

\bibitem{Ball:2004ye}
  P.~Ball and R.~Zwicky,  Phys.\ Rev.\ D {\bf 71}, 014015 (2005).

\bibitem{Bauer:2003mg}
  C.~W.~Bauer, D.~Pirjol and I.~W.~Stewart,  Phys.\ Rev.\ D {\bf 68},
  034021 (2003). 
  

\bibitem{Aubert:2003mm}
  B.~Aubert {\it et al.}  [BABAR Collaboration],
  Phys.\ Rev.\ Lett.\  {\bf 91}, 171802 (2003).

\bibitem{Aubert:2004xc}
  B.~Aubert {\it et al.}  [BABAR Collaboration],
  Phys.\ Rev.\ Lett.\  {\bf 93}, 231804 (2004).

\bibitem{Abe:2004ku}
  K.~Abe {\it et al.}  [BELLE Collaboration],
  arXiv:hep-ex/0408141.

\bibitem{rescattering}
  P.~Colangelo, F.~De Fazio and T.~N.~Pham,
  Phys.\ Lett.\ B {\bf 597}, 291 (2004);
  M.~Ladisa, V.~Laporta, G.~Nardulli and P.~Santorelli,
  Phys.\ Rev.\ D {\bf 70}, 114025 (2004);
  H.~Y.~Cheng, C.~K.~Chua and A.~Soni,
  Phys.\ Rev.\ D {\bf 71}, 014030 (2005).

\bibitem{Kagan:2004uw}
  A.~L.~Kagan,
  Phys.\ Lett.\ B {\bf 601}, 151 (2004).

\bibitem{Li}
  H.~n.~Li,  arXiv:hep-ph/0411305; H.~n.~Li and S.~Mishima,
  Phys.\ Rev.\ D {\bf 71}, 054025 (2005).
  


\end{thebibliography}
\end{document}